\documentclass[journal]{IEEEtran}

\usepackage{comment}
\usepackage{amsmath,amssymb,bm}
\usepackage{nicematrix,tikz}
\usepackage{tabularx}
\usepackage{graphicx}
\usepackage{booktabs} 
\usepackage{caption}
\usepackage[labelformat=simple]{subcaption}
\usepackage{xcolor}
\usepackage{cancel}
\usepackage{cite}
\usepackage[linesnumbered,ruled,vlined]{algorithm2e}
\usepackage{setspace}

\newtheorem{remark}{Remark}

\newcommand\norm[1]{\left\lVert#1\right\rVert}
\newcommand{\tran}{^{\mbox{\scriptsize T}}}
\newcommand{\herm}{^{\mbox{\scriptsize H}}}

\makeatletter
\def\thickhline{%
  \noalign{\ifnum0=`}\fi\hrule \@height \thickarrayrulewidth \futurelet
   \reserved@a\@xthickhline}
\def\@xthickhline{\ifx\reserved@a\thickhline
               \vskip\doublerulesep
               \vskip-\thickarrayrulewidth
             \fi
      \ifnum0=`{\fi}}
\makeatother

\newlength{\thickarrayrulewidth}
\ifCLASSINFOpdf
\else
\fi
\usepackage[acronym]{glossaries}

\makeglossaries

\newglossaryentry{latex}
{
        name=latex,
        description={Is a mark up language specially suited for 
scientific documents}
}

\newglossaryentry{maths}
{
        name=mathematics,
        description={Mathematics is what mathematicians do}
}

\newglossaryentry{formula}
{
        name=formula,
        description={A mathematical expression}
}

\newacronym{gf-noma}{GCF-NOMA}{Grant-Free non orthogonal multiple access}
\newacronym{mtds}{MTDs}{Machine type devices}
\begin{document}
\bstctlcite{IEEEexample:BSTcontrol}\color{black}{
\title{Joint Activity Detection and Channel Estimation for Clustered Massive Machine Type Communications}
\author{\IEEEauthorblockN{Leatile Marata, \textit{Member, IEEE}, Onel Luis Alcaraz López, \textit{Member, IEEE}, Andreas Hauptmann, \textit{Senior Member, IEEE}, Hamza Djelouat, \textit{Student Member, IEEE}, and Hirley Alves, \textit{Member, IEEE}
}
\thanks{Leatile Marata, Onel Luis Alcaraz López, Hamza Djelouat, and Hirley Alves are with Centre for Wireless Communications -- Radio Technologies, FI-90014, University of Oulu, Finland. e-mail: \{leatile.marata,onel.alcarazlopez,hamza.djelouat, hirley.alves\}@oulu.fi.  Andreas Hauptmann is with the Research Unit of Mathematical Sciences, FI-90014, University of Oulu, Finland, and also with the
Department of Computer Science, University College London, London
WC1E 6BT, U.K. e-mail: andreas.hauptmann@oulu.fi}
\thanks{This work is supported by the Academy of Finland (Grants n.319485, n.340171, n.346208 (6G Flagship), n.338408, n.353093). Leatile Marata's work was partly supported by the Riitta ja Jorma J. Takanen Foundation, the Finnish Foundation for Technology Promotion, and the Botswana International University of Science and Technology. The Finnish Foundation for Technology Promotion partly supported Onel López's work. The Tauno Tönning Foundation partly supported Hamza Djelouat's work. }}
\maketitle
\begin{abstract} Compressed sensing multi-user detection (CS-MUD) algorithms play a key role in optimizing grant-free (GF) non-orthogonal multiple access (NOMA) for massive machine-type communications (mMTC). However, current CS-MUD algorithms cannot be efficiently parallelized, \textcolor{black}{leading to} computationally expensive implementations of joint activity detection and channel estimation (JADCE) as the number of deployed machine-type devices (MTDs) increases. To address this, the present work proposes novel JADCE algorithms that can be applied in parallel for different clusters of MTDs by exploiting the structure of the pilot sequences. These are the approximation error method (AEM)-alternating direction method of multipliers (ADMM), and AEM-sparse Bayesian learning (SBL). Results presented in terms of the normalized mean square error and the probability of miss detection \textcolor{black}{show} comparable performance to the conventional algorithms. However, both AEM-ADMM and AEM-SBL algorithms have significantly reduced computational complexity and run times, thus, \textcolor{black}{facilitating} network scalability.  
\end{abstract}
%\begin{IEEEkeywords} Approximation error method, detection, grant-free, 
%machine-type communication, MIMO, sparse signal recovery.
%\end{IEEEkeywords}
\IEEEpeerreviewmaketitle

%% IN THIS PARAGRAPH YOU WILL DISCUSSIONS ARE CENTERED AROUND EVOLUTION OF WIRELESS COMMUNICATIONS AND HOW MTC IS DIFFERENT
\section{Introduction}
\label{IntroductionSect}
\IEEEPARstart{D}{etection}, channel estimation, and data decoding are fundamental operations performed by a receiver in a wireless communication network \cite{zhang2021overview,renfors2019signal,albreem2019massive}. However, the majority of algorithms designed for these operations in previous wireless communication systems, i.e., fourth-generation (4G) and earlier, were tailored exclusively for downlink human-type communications (HTC)\cite{spencer2004introduction,heath2016overview}. In a turn of events, the new communication standards, i.e., the fifth generation (5G) and beyond (5GB), natively support a new set of devices termed machine-type devices (MTDs), which perform various sensing tasks in the Internet of Things (IoT) paradigm\cite{lopez2022statistical,djelouat2021user,marata2021joint}. Notably, MTDs
are energy constrained, yet in some cases, they need to be deployed in remote areas where they cannot be readily charged. For this reason, MTDs are designed to save energy by  
only switching to active transmission mode after sensing data and remaining in sleep mode in the absence of data. This intermittent mode of operation reduces energy consumption while creating sporadic uplink traffic, which is unconventional to HTC. Since the modus operandi of the MTDs is incompatible with existing HTC devices, a verbatim implementation of existing receive algorithms in massive machine-type communication (mMTC) networks can degrade communication performance. Fortunately, the aforementioned problems can be jointly addressed by employing low-complexity transmission schemes, for which grant-free (GF) non-orthogonal multiple access (NOMA) plays a pivotal role.

%% THIS PARAGRAPH DISCUSSES GF-NOMA and its drawbacks 

\par GF-NOMA techniques have been proposed as low complexity transmission schemes for uncoordinated transmissions of the MTDs \cite{shahab2020grant,senel2018grant}. Under these schemes, active devices transmit their data without permission from the base station (BS), thus bypassing the signaling overheads that are associated with the handshaking/scheduling process  and consequently reducing communication overheads and access latency\cite{shahab2020grant,senel2018grant}. Nevertheless, the lack of scheduling and the inevitable use of non-orthogonal pilot sequences lead to increased collisions and multi-user interference (MUI). An inefficient use of GF-NOMA can be detrimental to the previously mentioned operations (detection, channel estimation, and data decoding), which is one of the major drawbacks of GF-NOMA. Ultimately, the performance of GF-NOMA  schemes relies on efficiently resolving both the collisions and MUI\cite{li2022joint}.

%% THIS PARAGRAPH FOCUSES ON SYSTEMATIC APPROACHES FOR OPTIMIZING GF-NOMA

\par The need for efficient GF-NOMA has motivated compressed sensing (CS) multi-user detection (MUD) for joint activity detection and channel estimation (JADCE) and/or unsourced random access (URA) \cite{fengler2021non}. The former is concerned with user identification and data decoding, while the latter is concerned with decoding the transmitted data \textcolor{black}{instead of} identifying the actual transmitting MTD\cite{shao2020cooperative}. As a result, JADCE is applicable in status update scenarios with different types of messages, e.g., when there are different MTDs for sensing ambient humidity, acidity, and temperature. Conversely, URA can be employed in scenarios where multiple MTDs transmit observations about a common physical phenomenon, such as temperature measurements in a smart factory, \textcolor{black}{to obtain average} information  about this phenomenon. \par By and large, both JADCE and URA rely on the fact that a relatively small number of MTDs are simultaneously active in a given coherence interval (CI) despite their massive numbers. For instance, future networks are expected to host up to 10 million MTDs per km$^{2}$, while only a small fraction of them will be active at the same time \cite{chen2020massive}. The identification of the active MTDs can naturally be posed as a CS-MUD problem, which can be computationally complex\cite{li2021compressed,liu2018sparse}. However, with the increasing deployments of IoT, some MTDs are bound to present similar characteristics and performance requirements. These similarities can naturally facilitate the formation of clusters\cite{di2020detection,zhu2022ofdm,jiang2022statistical,lopez2020ultra}. Basically, clusters of the MTDs can be formed according to the channel statistics (e.g., channel covariance matrix), performance requirements, traffic characteristics, or activation probabilities, among other things \cite{lopez2020ultra}. \textcolor{black}{Given these considerations, clustering the MTDs can help optimize resource allocation, thus making the network design more flexible and scalable. This can lead to simplifying some CS-MUD problems and facilitating GF-NOMA for a massive number of MTDs.}

% are needed to optimize GF-NOMA for a massive number of MTDs while leveraging the clusters. 
%% THIS PARAGRAPH DISCUSSES SOME SOLUTIONS TO THE DISCUSSED PROBLEMS 
\par There is a noteworthy research endeavor to develop scalable algorithms for CS-MUD to optimize GF-NOMA. Typically, these methods utilize the massive multiple-input multiple-output (mMIMO) technology, which enables distributed or parallel signal processing. In a quest to accommodate a massive number of MTDs, most of the works rely on  pilot data designed from fully non-orthogonal sequences \cite{ngo2015cell,rajoriya2022centralized,ganesan2021clustering,kim2022downlink,abdi2019optimization}. Even though non-orthogonality of the pilot sequences is crucial for serving a massive number of MTDs, it is possible to \textcolor{black}{devise} pilot sequences that can be grouped into finite orthogonal subspaces to capture/realize different clusters. Given the orthogonal subspaces, it is practical to implement CS-MUD algorithms in parallel while maintaining zero MUI across the clusters. Ultimately, this reduces the need for information exchange while the algorithms run in parallel. Equally important is to note that such pilot sequences have to be formed using non-independent identically distributed (i.i.d.) sequences such as Hadamard, Zadoff Chu, and Fourier matrices, \textcolor{black}{which are all} consistent with the \textcolor{black}{3GPP Release 17} standard \cite{3gpp_standard}. As noted by Liu \textit{et al.} \cite{liu2021decentralized}, using i.i.d. pilot sequences is impractical, and most CS-MUD algorithms designed thus far to work under this assumption face challenges in practical scenarios. It is therefore crucial to \textcolor{black}{develop} practical approaches that facilitate the efficient implementation of CS-MUD algorithms while considering the existence of clusters in mMTC. \textcolor{black}{Even though providing a good foundation for parallel JADCE design, to the best of our knowledge there are no works on JADCE algorithms that incorporate the orthogonal pilot subspaces for clustered MTC. This is precisely the aim of the present work.}  Some of the advantages of the proposed algorithms are: i) reduced MUI, ii)  efficient resource usage, iii) scalability, and iv) network design flexibility, all of which contribute to efficient signal recovery. To provide context, we present a brief literature survey of related works.

\subsection{Related Literature}\label{literatureReview} In the recent past, mMIMO-enabled mMTC  has become an active area of research. mMIMO is crucial for mMTC because it can increase spectral efficiency, data rates, and link reliability \cite{elhoushy2021cell}. There are some ongoing works to develop efficient CS-MUD algorithms for GF-NOMA using mMIMO. For example, He \textit{et al.} proposed a distributed detection algorithm  based on expectation propagation (EP) \textcolor{black}{in}\cite{he2021distributed} 
 to facilitate integration at the central processing unit (CPU). The proposed work also presented a performance analysis of the EP in a distributed cell-free (CF)-MIMO. However, although the proposal improves the detection performance, it also increases the computational complexity of the CPU. Similarly, Li \textit{et al.} \textcolor{black}{in} \cite{li2022covariance} proposed a covariance-based device activity detection algorithm that exploits orthogonal pilot sequences to reduce the MUI. Relative to existing works, their results showed improved performance for low signal-to-noise ratio (SNR) and short pilot lengths. On the other hand, Ganesan \textit{et al.} presented a maximum likelihood (ML) based device detection algorithm for CF-MIMO \textcolor{black}{in} \cite{ganesan2021clustering}. Their results demonstrated improved performance when using CF-MIMO as opposed to co-located MIMO. In general, ML algorithms proposed in \cite{ganesan2021clustering,li2022covariance} have high computational complexity.

\textcolor{black}{In a quest for lower computational complexity receivers, the approximate message passing (AMP) algorithm, first introduced by Donoho \textit{et al.} in \cite{donoho2009message} has been widely explored under different settings.} \textcolor{black}{For instance}, Bai \textit{et al.}  proposed a distributed AMP algorithm \textcolor{black}{in}\cite{bai2022activity}
based on the likelihood ratio and incorporated the structure of the state evolution. However, it should be noted that \textcolor{black}{most} AMP algorithms are only guaranteed to converge if the columns of the pilot matrix \textcolor{black}{follow the}  Gaussian distribution and are uncorrelated. \textcolor{black}{In general, this property of AMP algorithms limits their practical application}. \textcolor{black}{Motivated by this challenge, Rangan \textit{et al.} proposed the vector AMP (VAMP) in \cite{rangan2019vector} to guarantee convergence under broader structures of the columns of the pilot sequences. Similarly, Ma \textit{et al.} proposed the orthogonal AMP (OAMP)  in \cite{ma2017orthogonal}, which can converge under pilot sequences that are generated from partially orthogonal matrices. As a result, both VAMP and OAMP have been extensively explored for JADCE problems, such as in \cite{cheng2020orthogonal}, where Cheng \textit{et al.} proposed the OAMP as a solution for spatially and temporally correlated channels. Despite this, AMP, VAMP, and OAMP are designed to work in large-dimensional problems, hence performing poorly when applied in small-dimensional problems \cite{senel2018grant,cheng2020orthogonal}. } 

\textcolor{black}{Notably}, \cite{elhoushy2021cell,he2021distributed,li2022covariance,ganesan2021clustering,bai2022activity,cheng2020orthogonal} all assume that devices are synchronized, which inspired  Li \textit{et al.} \cite{li2022asynchronous} to propose an asynchronous device activity detection in CF-MIMO systems where communication between the BSs and the CPU is optimized. From the results of \cite{li2022asynchronous}, it is apparent that one of the bottlenecks of decentralized algorithms is the communication overhead incurred by increased signaling between the different sub-processors. In a similar vein, Chen \textit{et al.} \textcolor{black}{in} \cite{chen2020structured} proposed a structured massive access for CF-MIMO using the per group and the IB-K-means clustering algorithms. Their work showed an improved spectral efficiency of the proposed pilot assignment strategies, outperforming conventional pilot assignments. In addition, Figueredo \textit{et al.} \textcolor{black}{in}\cite{de2018application} presented a feasibility study for improving system capacity by clustering the MTDs such that they can share the same time-frequency resource blocks. Iimori \textit{et al.} \textcolor{black}{in} \cite{iimori2022joint} proposed a bi-linear message passing algorithm that efficiently detects clusters of devices by leveraging the sparsity in the sub-arrays of extra-large MIMO.\par
Despite the potential benefit of capturing the clusters of MTDs using orthogonal pilot subspaces to facilitate efficient parallel implementation  of JADCE algorithms, this has not been explored in the literature.  Motivated by the work \cite{marata2022joint}, where Marata \textit{et al.} proposed some pilot design strategies to enable the amicable coexistence of different services, we present novel CS-MUD algorithms that exploit the pilot structure of the clusters of MTDs in mMTC scenarios.
\subsection{Contributions}
We consider an mMIMO network serving heterogeneous clusters of MTDs\footnote{Here, heterogeneity refers to differences in characteristics and performance requirements of the MTDs.}. By capturing the heterogeneous characteristics of the MTDs using orthogonal pilot subspaces, we present a JADCE problem and solve it using parallel algorithms. Notice that, the present work departs from works such as \cite{elhoushy2021cell,he2021distributed,li2022covariance,ganesan2021clustering,bai2022activity}, where some iteration steps are exchanged even in parallel implementations of the CS-MUD algorithms. \textcolor{black}{Our main contributions are} as follows:  
\begin{itemize}
    \item We formulate the JADCE problem based on pilot subspaces, i.e., where the massive non-orthogonal pilot sequences of the MTDs of each cluster are computed from orthogonal subspaces. Some of the main advantages of the proposed formulation are the parallel implementation of the sparse signal recovery (SSR) algorithms, network design flexibility, and scalability. 
     \item We propose data-driven algorithms that utilize the approximation error method (AEM) established in the field of inverse problems to account for errors in the sensing matrix and likelihood function \cite{mozumder2021model,lunz2021learned,kaipio2006statistical}. Herein, AEM is used to account for the mismatch between the ideal measurement and the measurement used to perform JADCE in each cluster. First, we propose the AEM-alternating direction method of multipliers (ADMM), which leverages the learned statistics of the mismatch to perform iterative soft thresholding. Second, we present the AEM-sparse Bayesian learning (SBL) algorithm which exploits a corrected likelihood function within the Bayesian framework. AEM-ADMM does not take the prior distribution into consideration and is applicable for scenarios without distributions of the parameters. On the other hand, AEM-SBL relies on statistical distributions, thus utilizing more information to improve the JADCE performance.  
\item We compare the proposed JADCE framework with the conventional approaches, which are applied without clustering, and numerically quantify their performance. Specifically, we show that our proposed algorithms achieve comparable channel estimation accuracy and detection capabilities to their classical counterparts while benefiting from reduced run-time.  
\end{itemize}
\subsection{Organization and Notation}
The remainder of this paper is organized as follows. \textcolor{black}Section \ref{formulate} {introduces the system model.  Section~\ref{Decentralize} presents} the cluster-based device activity detection problem. In Section \ref{detections}, we \textcolor{black}{propose solutions} to this problem, while Section \ref{results} presents the results and discussions. Lastly, in Section~\ref{Conclude}, we conclude the paper and discuss some future research directions.
\par \textbf{Notation:} \textcolor{black}{Boldface lowercase} and uppercase letters denote column vectors and matrices, respectively. Moreover, $\mathbf{a}_i$ and $a_{i,j}$ are the $i$-th column and the element in the $i$-row, $j$-th column of matrix $\mathbf{A}$, respectively, while $a_i$ is the $i$-the element of vector $\mathbf{a}$. The superscripts $(\cdot)^*$, $(\cdot)\tran$, and $(\cdot)\herm$ denote the conjugate, transpose, and conjugate transpose operations, respectively. For both matrices and vectors, the hat notation indicates an estimate, e.g., $\hat{x}$ is the estimate of $x$. Additionally, $\mathbb{C}$ and $\mathbb{R}$ refer to complex and real domains, respectively. We denote the circularly symmetric complex Gaussian distribution with mean $\mathbf{a}$ and covariance $\mathbf{B}$ by $\mathcal{CN}(\mathbf{a},\mathbf{B})$, while $\mathbb{E}\{\cdot\}$ and $\mathbb{V}\{\cdot\}$ are the expectation and covariance operators, respectively. \textcolor{black}{Additionally,  $\mathcal{U}(a,b)$ denotes the uniform distribution with bounds $a$ and $b$.} The $\mathrm{diag}\{a_{1},a_{2},\cdots,a_{n}\}$ creates a diagonal matrix whose main diagonal terms are $a_{1},a_{2},\cdots,a_{n}$. Finally, $\norm{\cdot}_{F}$, $\norm{.}_{p}$ and \textcolor{black}{$\lVert \cdot \rVert_{n,p}$} denote the Frobenius norm, $\ell_p$ norm and \textcolor{black}{mixed $n, p$ norm}, respectively, while the probability distribution of random variables is defined as $\mathcal{P}(\cdot)$, while $\mathcal{P}(\cdot|\cdot)$ is the conditional probability. 
\section{System model}\label{formulate}
\begin{figure}
    \centering
    \includegraphics[scale=0.4]{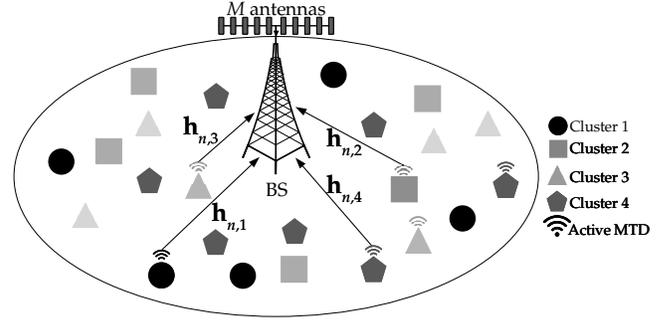}
    \caption{An mMTC scenario where an $M$-antenna BS serves $N$ MTDs grouped into $G$ clusters, among which an average total of $\epsilon N$ MTDs are active.}
    \label{figMTC}
\end{figure}
We consider the uplink massive MIMO scenario depicted by Fig~\ref{figMTC}, where a BS equipped with a set of $M$ antennas, i.e., $\mathcal{M} =\{1,\cdots,M\}$ serves a set $\mathcal{N} = \{1, \cdots,N\}$ of stationary MTDs. Among these, a subset $\mathcal{K}\subset\mathcal{N}$ of cardinality $K=|\mathcal{K}|$ is active and unknown to the BS. The MTDs are assumed to be active with a probability $\epsilon\ll1$, thus the average number of active devices in the network  in a given CI is $\epsilon N$. Moreover, $\mathcal{N}$ is sub grouped into a set of clusters sub-indexed by $\mathcal{G}=\{1,\cdots,G\}$, where $G\leq N$ and each MTD exclusively belongs to a cluster $\mathcal{C}_{g} \subseteq  \mathcal{N}$, $g\in\mathcal{G}$. The cardinality  of the $g$-th cluster is defined as  $N_g$, such that $\sum_{g=1}^{G}N_{g} = N$. \par We also assume quasi-static block fading channels, \textcolor{black}{such that} channels remain unchanged during each CI of $T$ symbols and change independently \textcolor{black}{between CIs}. We assume that only the large-scale channel state information (CSI), i.e., the path loss information, is available at the BS and not the instantaneous CSI. \textcolor{black}{The uplink channel between the $n$-th MTD in the $g$-th cluster and the BS in a given CI is defined as $\mathbf{h}_{n,g} \sim \mathcal{CN}(\mathbf{0}, \mathbf{Q}_{n,g}) \in \mathbb{C}^{M\times 1}$, where $\mathbf{Q}_{n,g}$ is the channel covariance matrix.} Let $\alpha_{n,g}$ denote the activation status of device $n$ in the $g$-th cluster as
\begin{equation}
  \alpha_{n,g}\!=\! 
  \begin{cases} 
      1,~\text{if device $n$ in the $g$-th cluster is active}\\
      0,~\text{otherwise}\\
   \end{cases},  
   \label{actFuction}
\end{equation}
hence, the overall network activity indicator is given by the vector ${\bm{\alpha}\!=\! [\bm{\alpha}_{1}\tran,\bm{\alpha}_{2}\tran,\cdots,\bm{\alpha}_{G}\tran]\tran}\in \{0,1\}^{N\times 1}$, where $\bm{\alpha}_{g}\!=\![\alpha_{g,1},\alpha_{g,2},\cdots,\alpha_{g,N_g}]\tran\in \{0,1\}^{N_g\times 1}$, $\forall g$. The first phase of each CI (defined as the first $L$ symbols) corresponds to the metadata processing block, i.e., where JADCE is carried out, while the other $T-L$ symbols are used to  convey the intended message (payload). Each
MTD is therefore pre-allocated a pilot sequence $\mathbf{s}_{n,g}\in \mathbb{C}^{L\times 1}$ that is known to both the BS and the MTD for JADCE\footnote{Given that the pilot sequences are known \emph{a priori} by the BS, it is possible to optimize their structure to improve the receive algorithms. To that end, we also present a combinatorial problem in Section~\ref{Decentralize} that can be solved at the BS to improve the detection capabilities.}. However, due to the massiveness of the MTDs, the length of each pilot sequence is usually much smaller than the total number of devices, i.e., $L\ll N$. Consequently, the BS must pre-allocate non-orthogonal pilot sequences to all the MTDs and employ CS-MUD to solve the JADCE problem based on the signal received during the training phase, which is given by {\color{black} \begin{align}\label{mtc}
\mathbf{Y}&=\sum_{\forall n,g}\alpha_{n,g}\sqrt{p_{n,g}}\mathbf{s}_{n,g}\mathbf{h}_{n,g}\tran+\mathbf{W}
  \nonumber \\
  &=\sum_{g=1}^{G}\mathbf{S}_{g}\mathbf{X}_{g}+\mathbf{W}
   \nonumber \\
  &
  =\mathbf{S}\mathbf{X}+\mathbf{W},
\end{align}}
where ${\mathbf{S}_g\!=\![\mathbf{s}_{n,g},\cdots,\mathbf{s}_{N_g,g}]}$, ${\mathbf{X}_g \!=\! [\mathbf{x}_{1,g},\cdots,\mathbf{x}_{N_g,g}]\tran}$ $\forall g$, $\mathbf{S}\!=\![\mathbf{S}_1\tran,\cdots,\mathbf{S}_G\tran]\tran$, and $\mathbf{X}\!=\![\mathbf{X}_1\tran,\cdots,\mathbf{X}_G\tran]\tran$. Meanwhile, ${\mathbf{W} \in \mathbb{C}^{L\times M}}$ is the receiver noise, whose columns are i.i.d. as $\{\mathbf{w}_m\}\sim \mathcal{CN}(\bm{0},\sigma^2\mathbf{I})\in \mathbb{C}^{L \times 1}$, while ${\mathbf{x}_{n,g} =\sqrt{p_{n,g}}  \alpha_{n,g}\mathbf{h}_{n,g}\in\mathbb{C}^{M\times 1}}$ and $p_{n,g}$  are the effective row-sparse channel vector and the transmit power of the $n$-th device in the $g$-th cluster, respectively. \par  From \eqref{mtc}, it is evident that if $\{\mathbf{S}_{g}\}$, $\forall g$, are created from orthogonal subspaces, it is possible to completely decentralize the JADCE process by performing the detection of each cluster separately, thus reducing the complexity of the problem. To this end, we subsequently discuss the formation of $\mathbf{S}_{g}$, $g\!=\!1,\cdots, G$, using orthogonal subspaces, which allows fully decentralized detection and channel estimation.

%% WE START THE DISCUSSION ABOUT DECENTRALIZED DETECTION  
\section{Decentralized detection as a CS problem}
\label{Decentralize}
\textcolor{black}{To achieve decentralization, we assume that the pilot sequences $\{\mathbf{S}_{g} \in \mathbb{C}^{L\times N_g}\}$ associated with each cluster are strictly generated as linear combinations of orthogonal basis matrices. Note that the columns of an orthogonal basis matrix are mutually orthogonal (perpendicular) vectors \cite{strang1993introduction}. For this reason, orthogonal basis matrices provide a good foundation for handling structured pilot design. Examples of orthogonal basis matrices for designing pilot sequences in MTC include the identity matrix, Hadamard matrix, and Fourier matrices, among others \cite{bjornson2017massive}. However, the identity matrix lacks the diversity required for sensing matrices \cite{choi2017compressed}. Although both the Fourier matrix and the Hadamard matrix can serve as orthogonal basis matrices, we adopted the Hadamard matrix due to its appealing computational and storage properties. Let $\mathbf{B} \in \mathbb{C}^{L \times L}$ be such a square matrix whose columns can be partitioned into $G$ different orthogonal matrices, i.e., $\mathbf{B}\!=\![\mathbf{B}_{1},\mathbf{B}_{2},\cdots,\mathbf{B}_{G}] \in \mathbb{C}^{L \times L}$, $\norm{\mathbf{B}_{i}\herm \mathbf{B}_{j}}_{F}\!=\!0$, $\forall i \neq j$. Consequently, the pilot sequences $\mathbf{S}_g$ of the $g$-th cluster are generated from $\mathbf{B}_{g}\in \mathbb{C}^{L \times \kappa_{g}}$, where $\kappa_g$ denotes the number of columns of $\mathbf{B}_{g}$ used for the $g$-th cluster, such that $\sum_{\forall g}\kappa_g \leq L$.} \par As discussed in  Section~\ref{IntroductionSect}, the number of MTDs is generally massive, and thus $N_g \geq L$ is a valid assumption. It is therefore computationally prohibitive to generate large $\mathbf{B}_{g}$. Furthermore, the CI is always finite, thus, infeasible to allocate mutually orthogonal pilot sequences to the MTDs belonging to the same cluster. Owing to the condition $N_g \geq L$, it is necessary to generate each $\mathbf{S}_g$ while guaranteeing signal recovery for each cluster, which is a fundamental problem in sensing (measurement) matrix design \cite{zhang2023physics}. From CS perspectives, the matrix $\mathbf{S}_g$ guarantees signal recovery if it satisfies the restricted isometric property (RIP), formally stated  as 
\begin{equation}
   (1\!-\!\delta_{\epsilon})\norm{\mathbf{X}_{g}}_{F}^{2}\leq \norm{\mathbf{S}_{g}\mathbf{X}_{g}}_{F}^{2}\leq (1\!+\!\delta_{\epsilon})\norm{\mathbf{X}_{g}}_{F}^{2},\forall \mathbf{X}_g,
\end{equation}
such that $\norm{\mathbf{X}_g}_{2,0}\leq K_g$, where $K_g$ is the average number of active devices in a given cluster and $\delta_{\epsilon}>0$ is the restricted isometric constant  \cite{eldar2012compressed}. 
\begin{remark}
The RIP can be interpreted as the ability of the matrix $\mathbf{S}_g$ to map $\mathbf{X}_{g}$ into the measurement space while maintaining the separation between the different samples of $\mathbf{X}_{g}$. This makes it possible to recover different samples of $\mathbf{X}_{g}$ without ambiguity.
\label{remarkRIP}
\end{remark}
\par From the Remark~\ref{remarkRIP} and without any loss of generality, note that creating pilot sequences for two different MTDs within a cluster, i.e., $\mathbf{s}_{i,g}$ and $\mathbf{s}_{j,g}$, with $i \neq j$, involves maximizing the minimum distance between two distinct pilot sequences, i.e.,
% \begin{equation}
%   \begin{array}{ll}
% \underset{\mathbf{S}_g}{\operatorname*{maximize}} \quad & \underset{1\leq i < j \leq |\mathcal{C}_g|}{\operatorname*{min}}d(\mathbf{B}_{g}\mathbf{z}_{i},\mathbf{B}_{g}\mathbf{z}_{j}) \\
% \textrm{subject to} \quad & ~\norm{\mathbf{z}_i}_{0} = \norm{\mathbf{z}_j}_{0},
% \\
% & ~\textcolor{black}{\mathbf{s}_{k,g}= \mathbf{B}_{g}\mathbf{z}_k},
% \end{array}
% \label{zOpt} 
% \end{equation}

\begin{subequations}\label{zOpt}
    \begin{alignat}{2}
&\underset{\mathbf{S}_g}{\mathrm{maximize}} \quad &&\underset{1\leq i < j \leq |\mathcal{C}_g|}{\mathrm{min}} \ d\left(\mathbf{B}_g\mathbf{z}_i,\mathbf{B}_g\mathbf{z}_j\right) \label{P1a}\\
&\text{subject to} \ && \lVert \mathbf{z}_i \rVert_0 = \lVert \mathbf{z}_j \rVert_0,  \label{P1b}\\
& \ && \textcolor{black}{s_{k,g} = \mathbf{B}_g\mathbf{z}_k, }\label{P1c}
    \end{alignat}
\end{subequations}

\noindent where $d(\cdot,\cdot)$ is a generic distance measure between the two entries, while $\mathbf{z}_{k} \in\mathbb{C}^{\kappa_g\times 1}$, $k\in \mathcal{C}_g$ is a vector containing the random combining weights with an optimized cardinality. Observe that $\mathbf{z}_k$ must be sparse to guarantee good detectability, thus ${\mathbf{s}_{k,g} = \mathbf{B}_{g}\mathbf{z}_{k}}$, $k\in \mathcal{C}_g$, whereas the equality constraint \eqref{P1b} ensures fairness. {\color{black} \begin{figure*}[t!]
 {\color{black}\begin{align}
     \mathbf{B}\!=\!
\begin{pNiceMatrix}
  1+ j&1+ j&1+ j&1+ j&1+ j&1+ j&1+ j&1+ j\\  
 1+ j&-1- j&1+ j&-1- j&1+ j&-1-j&1+j&-1- j\\   
1+ j&1+j&-1- j&-1- j&1+ j&1+ j&-1- j&-1- j\\
     1+ j&-1- j&-1- j&1+ j&1+ j&-1- j&-1- j&1+ j\\
     1+ j&1+ j&1+ j&1+ j&-1- j&-1- j&-1-j&-1- j\\
     1+ j&-1- j&1+ j&-1-j&-1- j&1+ j&-1- j&1+ j\\
     1+ j&1+ j&-1- j&-1- j&-1- j&-1- j&1+ j&1+ j\\
     1+ j&-1- j&-1- j&1+ j&-1-j&1+j&1+ j&-1-j 
     \CodeAfter
  \UnderBrace[shorten,yshift=1.5mm]{last-1}{last-3}{\mathbf{B}_1}
  \UnderBrace[shorten,yshift=1.5mm]{last-4}{last-6}{\mathbf{B}_2}
  \UnderBrace[shorten,yshift=1.5mm]{last-7}{last-last}{\mathbf{B}_3}
\end{pNiceMatrix},\\\nonumber\\\nonumber\\
     \bottomrule\nonumber
     \end{align}}
 \end{figure*}} \par Due to its combinatorial nature, the problem in \eqref{zOpt} is NP-hard. To provide its approximate solution, we exploit the procedure discussed in Section IV-B of \cite{marata2022joint}. This will yield $\{\mathbf{S}_{g} \in \mathbb{C}^{L \times N_g}\}$ for each cluster, resulting in a  concatenated pilot matrix $\mathbf{S}\!=\![\mathbf{S}_{1},\mathbf{S}_{2},\cdots,\mathbf{S}_{G}] \in \mathbb{C}^{L \times N}$ of all the devices in the network\footnote{To simplify the presentation, the matrix concatenation is assumed to follow the order of $g=1,\cdots, G$.}. This essentially leads to pilot sequences that are orthogonal to one another for different clusters\footnote{\textcolor{black}{The pilot sequences can be fixed for longer periods of time to reduce the computational complexity associated with indicating the indices of the pilot sequences for each MTD. In this case, MTDs can store a list of pilot sequences in their local memory, and the BS can indicate the $N$ indices using $\log_{2}N$ bits. Alternatively, the BS can indicate the sequence using $L\log_2d$ bits, for a modulation scheme employing $d$ symbols.}}, i.e., $\lvert \mathbf{S}_{i}\herm \mathbf{S}_{j}\rvert_{F}\!=\!0$, $\forall i \neq j $.
% In Fig.~\ref{corelationFigs}, we present the results that illustrate the correlation of the pilot sequences for different $G$, $N$, and $L$. 
% \begin{figure}
% \centering\includegraphics[scale = 0.45]{resultsFigs/corelationPlotView.eps}
%     \caption{Correlation of the pilot sequences as a function of $N$ and $G$.}
%     \label{corelationFigs}
% \end{figure}
% {\color{black} To clarify the generation of each $g$-th cluster, we provided a toy example below. Assume that 
% \begin{equation}
% \mathbf{B}\!=\!
% \begin{bmatrix}
% \underbrace{\alpha_1^1{g}_{11},{g}_{12},\ldots,{g}_{1M}\\
% \alpha_1^2 {g}_{11},{g}_{12},\ldots,{g}_{1M}}_{ggg}
% \end{bmatrix},
% \label{newSparse}
% \end{equation}}
{\color{black} To clarify the generation of the pilot sequences for each $g$-th cluster, we provide a toy example below. Assume that $\mathbf{B}$ is given by equation (5) at the top of the next page,
and thus $L = 8$, $\kappa_1= 3$, $\kappa_2 = 3$, and $\kappa_3 = 2$. Focusing now on cluster 2, for which  $\mathbf{S}_2$ is formed from a linear combination of the columns of $\mathbf{B}_2\in \mathbb{C}^{L\times 3}$. For instance, let $\mathbf{z}_{1} = [1 ~0 ~ 1 ]\tran$  and $\mathbf{z}_{2} = [0 ~1~ 1 ]\tran$, $\mathbf{\bar{s}}_{1,2}$ and $\mathbf{\bar{s}}_{2,2}$ be vectors generated by 
\begin{align}
        \mathbf{\bar{s}}_{1,2} &\!=\! \mathbf{B}_{2}\mathbf{z}_{1}\!=\! [2 \!+\! 2j,\!-\!2\! +\! 2j,0,0,0,0,\!-\!2\! -\!2j,2 \!+\! 2j]\tran, \\ 
%\end{align}
%\begin{align}
     \mathbf{\bar{s}}_{2,2} &\!=\! \mathbf{B}_{2}\mathbf{z}_{2}\!=\! [2 \!+\! 2j,  0, 2 \!+\! 2j,0,\!-\!2 -\!2j,0,\!-\!2\! -\! 2j,0 ]\tran, 
\end{align}
which are normalized, i.e., $\mathbf{s}_{1,2} = \frac{\mathbf{\bar{s}_{1,2}}}{\lVert \mathbf{\bar{s}_{1,2} \rVert}}_{2}$ and $\mathbf{s}_{2,2} = \frac{\mathbf{\bar{s}_{2,2}}}{\lVert \mathbf{\bar{s}_{2,2} \rVert}}_{2}$ to form  $\mathbf{S}_2 = [\mathbf{s}_{1,2}, \mathbf{s}_{2,2},\cdots,\mathbf{s}_{N_g,2} ]$}. Note that the resulting matrices $\{\mathbf{S}_g\}$ can have highly correlated columns, which can restrict the applicability of certain state-of-the-art CS-MUD algorithms that rely on the AMP framework\cite{senel2018grant,chen2018sparse}. {\color{black}\begin{figure}[t!]
     \centering     \includegraphics[width=0.95\linewidth]{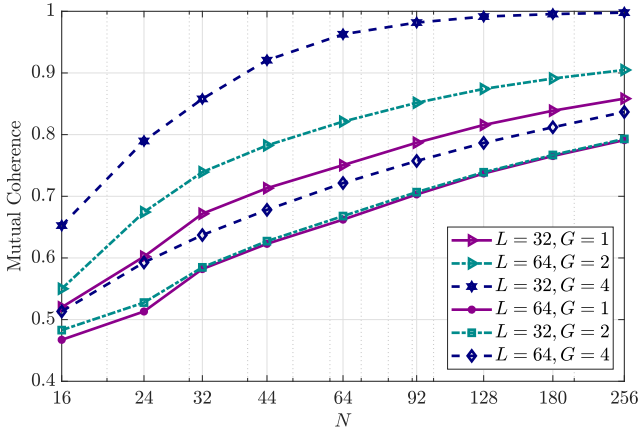}
     \caption{\textcolor{black}{Mutual coherence as a function of $L$, $G$, and $N$.}}
     \label{corelationFigs}
 \end{figure} \textcolor{black}{To exemplify,}  we illustrate in Fig.~\ref{corelationFigs} the maximum correlation between two different columns, i.e., the mutual coherence as a function of the number of MTDs ($N$) and clusters $G$. From the results, a relatively large number of clusters and a relatively small $L$ can increase the mutual coherence. This kind of result can therefore provide a guide on the proper choice of $L$ and $G$ for a given number of MTDs.}
\subsection{Device Activity Detection as a CS problem}From definition \eqref{actFuction} and the fact that traffic from MTDs is normally sporadic, the recovery of $\mathbf{X}$ constitutes a CS-MUD problem that can be solved using SSR concepts. Specifically, the BS has to identify the active devices from the compressed measurement $\mathbf{Y}$ with the knowledge that $\mathbf{X}$ is row-sparse. We can therefore define a generic inverse operation $f$ of the form $\hat{\mathbf{X}}\!=\!f(\mathbf{Y},\mathbf{S})$, that maps the measurement into the effective channel space. Furthermore, due to the orthogonality of the subspaces, the inverse operation can further be cast as  $\hat{\mathbf{X}}_g\!=\!f(\mathbf{Y}_g,\mathbf{S}_g)$ if $\mathbf{Y}_g$ is precisely known. Here, it is apparent that  $\hat{\mathbf{X}}$ would be a solution from a centralized problem such as those in \cite{wei2018joint,senel2018grant}, while $\hat{\mathbf{X}}_g$ would yield a solution for each cluster. Such an inverse operation can be formulated by exploiting Bayesian theory and/or a relaxed convex optimization framework.\par For the Bayesian formulation,  the sparsity promoting distribution of the effective channel $\mathbf{X}$  at each group level and within each group is modeled using the Bernoulli-Gaussian mixture distribution as \cite{iimori2022joint}
% \begin{equation}
%   \mathcal{P}(\mathbf{X})\!=\!\prod_{n,g}\Bigl((1-\epsilon)\delta(\mathbf{x}_{n,g}) + \epsilon\mathcal{CN}(\mathbf{x}_{n,g};\mathbf{0},\beta_{n,g}\mathbf{I}) \Bigr),
%     \label{prior}
% \end{equation}
\begin{equation}
  \mathcal{P}(\mathbf{X})\!=\!\prod_{n,g}\Bigl((1-\epsilon)\delta(\mathbf{x}_{n,g}) + \epsilon\mathcal{CN}(\mathbf{x}_{n,g};\mathbf{0},\textcolor{black}{\mathbf{\Tilde{Q}}_{n,g}}) \Bigr),
    \label{prior}
\end{equation}
where $\delta(\cdot)$ is the Dirac delta function imposing $\mathbf{x}_{n,g} = \bm{0}$ with a probability of $1-\epsilon$, \textcolor{black}{while $\mathbf{\Tilde{Q}}_{n,g} = p_{n,g}\mathbf{Q}_{n,g}$ is the effective covariance matrix.} 
From a mathematical perspective, precise knowledge of the linear problem \eqref{mtc} and \eqref{prior} makes it possible to compute a Bayes optimal $f$ to recover $\hat{\mathbf{X}}$  using the  maximum a posterior (MAP) estimate. As a consequence, the optimal JADCE algorithm \textcolor{black}{chooses} a pair of $\bm{\alpha}$ and $\mathbf{X}$ using
\begin{equation} \bm{\hat{\alpha}}\!=\!\operatorname*{argmax}_{\bm{\alpha}}\int \mathcal{P}(\mathbf{X},\bm{\alpha}|\mathbf{Y})d\mathbf{X},\label{mapEst1}
\end{equation} 
\begin{equation} \mathbf{\hat{X}}\!=\!\operatorname*{argmax}_{\mathbf{X}}\int \mathcal{P}(\mathbf{X},\bm{\alpha}|\mathbf{Y})d\bm{\alpha}\label{mapEst2}
\end{equation}
from the joint posterior distribution
\begin{equation}  
  \mathcal{P}(\mathbf{X},\bm{\alpha}|\mathbf{Y}) \propto \mathcal{P}(\mathbf{Y}|\mathbf{X},\bm{\alpha})\mathcal{P}(\mathbf{X}|\bm{\alpha})\mathcal{P}(\bm{\alpha}). 
  \label{posteriorCentralized}
  \end{equation}
  However, \textcolor{black}{the detector} that computes \eqref{mapEst1} and \eqref{mapEst2} is not practically implementable in a receiver. Firstly, there is a lack of precise information about the activation probabilities of the MTDs and that makes it difficult to formulate the prior distribution $\mathcal{P}(\mathbf{X})$ accurately. Secondly, even if the prior distribution can be accurately computed, the marginalization of \eqref{posteriorCentralized} requires prohibitively high dimensional integrals/summations with respect to a large number of variables in mMTC scenarios. \textcolor{black}{Hence}, most solutions to \eqref{mapEst1} and \eqref{mapEst2} are sought using alternative approaches. \par A common approach that relaxes the complex marginalization is to approximate the joint posterior distribution using belief propagation \cite{di2022joint}. Such an approach yields an efficient solution via the sum-product algorithm. Another approach involves the approximation of the posterior distribution using mean field techniques, under which the solution is found through variational message passing \cite{liu2019new,zhang2021unifying}. Alternatively, the solution can be sought from the relaxed convex optimization formulation
\begin{equation}
 \operatorname*{minimize}_{\mathbf{X}}
 \frac{1}{2}\norm{\mathbf{Y}-\mathbf{S}\mathbf{X}}_{F}^2\!+\!\lambda\norm{\mathbf{X}}_{2,1},
    \label{inrodOptimization}
\end{equation}
where $\lambda \in \mathbb{R}^{+}$ serves as a penalty term that trades off measurement fidelity and  sparsity structure  captured by the mixed $\ell_{2,1}$ norm. Nevertheless, observe that \eqref{inrodOptimization} can also be computationally burdensome for large-scale problems  and its effectiveness depends on properly choosing the penalty term. This is evidently sub-optimal compared to its Bayesian-based counterparts. However, it can achieve reasonable results and there have been some research efforts directed towards executing its solution in parallel, e.g., \cite{boyd2011distributed,djelouat2021spatial}. As alluded to earlier, such parallel implementations cannot reap the full benefits of the proposed pilot-based cluster model because they rely on the ability to decompose the objective function as opposed to the possible isolation of the clusters of the MTDs. In the end, their parallel implementations require the exchange of the updates of the Lagrange multipliers. We subsequently present the AEM-inspired SSR solutions that perform JADCE while considering the existence of clusters.
\section{Approximation error method inspired sparse recovery}
\label{detections}
Considering that \textcolor{black}{problem \eqref{mtc} is formulated such} that the pilot sequences $\mathbf{S}_i$ and $\mathbf{S}_j$ of $\mathcal{C}_i$ and $\mathcal{C}_j$, $i\neq j$, respectively, are orthogonal to each other, then the JADCE is broken down into smaller problems, one for each cluster. For example, the signal processing at the $g$-th cluster can depart from
\begin{equation}\Tilde{\mathbf{X}}_{g}\!=\!\mathbf{S}_{g}\herm\mathbf{Y}. 
    \label{decorelated}
\end{equation}
Interestingly, if the devices follow URA or have correlated activity, their joint activity can be estimated using ${ \lVert \Tilde{\mathbf{X}}_g \rVert_{F} \geq \zeta}$, where $\zeta$ should be greater but relatively close to $\frac{1}{\sqrt{L}}\sigma$ for good performance. The ability to handle both URA and correlated activity provides the much needed flexibility in pilot sequence allocation. For example, there is a  flexibility to allocate larger basis matrices $\mathbf{B}_g$ to devices that require more resources, such as those used in URLLC systems, as discussed by Lopez \textit{et al.} in \cite{lopez2020ultra}.  \par Observe that the decorrelation step  \eqref{decorelated} isolates the different clusters. However, $\Tilde{\mathbf{X}}_{g}$ resulted from a backward projection of the measurement into the solution space, thus, relying on this statistic for the JADCE can lead to sub-optimal performance. This is because the majority of existing algorithms are specifically designed to utilize the low-dimensional measurements captured by $\mathbf{Y}$ in order to compute $\hat{\mathbf{X}}$. Evidently, $\Tilde{\mathbf{X}}_{g}$ leads to the loss of crucial structures in the signal that is used for the JADCE at each cluster.  
\begin{remark}
Observe from \eqref{decorelated} that most CS-MUD algorithms that have the correlation step $\mathbf{S}_{j}\herm\mathbf{Y}$, $j \in \mathcal{G}$ among their iterative steps can provide a naive solution of the cluster based JADCE. 
\label{remark1}
\end{remark} Following the Remark~\ref{remark1}, we exemplify the implementation of the naive solution via simultaneous orthogonal matching pursuit (SOMP) \cite{cai2011orthogonal}. In order to differentiate it from the traditional implementation of SOMP, we refer to the cluster-based implementation as cluster-based SOMP (CB-SOMP), which is outlined in \textbf{Algorithm}~\ref{algoOMP}. In all the algorithms, $\Delta$ is the error tolerance level used in the stopping criteria. 
 
\begin{algorithm}[!t]
  \KwIn{$\mathbf{Y}$, $\Delta$}
 % \KwOut{$\hat{\bm{X}}^t$}

  $\hat{\mathbf{X}}_{g}^{(0)} = \bm{0}$\\$\mathbf{R}_{g}^{(0)}=\mathbf{Y}$, $t= 0$, $\mathcal{H}_{g}^{(0)} = \emptyset $\\\
    \Repeat{$   \frac{\norm{\mathbf{X}^{(t)}-\mathbf{X}^{(t-1)}}_{F}}{\norm{\mathbf{X}^{(t)}}_{F}} < \Delta$}{%
      %\label{stepB}

       $\mathbf{D}^{(t)} = \mathbf{S}_{g}\herm \mathbf{R}_{g}^{(t)}$\\
     
        $j^{(t)} = \displaystyle\text{argmax}_{j}   \left\{\frac{\norm{\mathbf{d}_{j}^{(t)}}_{1}}{\norm{\mathbf{s}_j}_{2}} \right\} $\
         % $T^{[t]} = T^{i-1}\cup j^{t} \bm{R}^{[t-1]}$\
          
          $\mathcal{H}^{(t)} = \mathcal{H}^{(t)}\cup j^{(t)} $\\
      % $j^{[t]} = \text{argmax}_{j}\frac{|g^{[t]}|}{\norm{\bm{A}_j}_{2}}$\\
           $\mathbf{X}_{[\mathcal{H}^{(t)}]}^{(t)} =\mathbf{S}_{[\mathcal{H}^{(t)}]}^{\dagger} \mathbf{Y} $\\
        $\mathbf{R}^{(t)} = \mathbf{Y} - \mathbf{S}\mathbf{X}^{(t)}$\
        
       $t = t + 1$\
       
    }
     \KwOut{$\mathbf{\hat{X}_{g}} = \mathbf{X}^{(t)}$}
  \caption{CB-SOMP,  $\forall g \in \mathcal{G}$}
  \label{algoOMP}
\end{algorithm}
\par As previously mentioned, the backward projection can be detrimental to the receiver's performance. A classical solution to map $\Tilde{\mathbf{X}}_{g}$ back into the measurement space of each of the clusters is by using \begin{equation}
    \mathbf{\hat{Y}_{g}} =(\mathbf{S}_{g}\mathbf{S}_{g}\herm)^{-1}\mathbf{S}_{g}\mathbf{\tilde{X}}_g. 
    \label{pseudoInv}
\end{equation} Meanwhile, similar to \eqref{posteriorCentralized}, the resulting reduced MAP problem has the joint posterior distribution
 \begin{equation} 
 \mathcal{P}(\mathbf{X}_{g},\bm{\alpha}_{g}|\hat{\mathbf{Y}}_{g})\!\propto\!\mathcal{P}(\hat{\mathbf{Y}}_{g}|\mathbf{X}_{g},\bm{\alpha}_{g})\mathcal{P}(\mathbf{X}_{g}|\bm{\alpha}_{g})\mathcal{P}(\bm{\alpha}_{g}). 
  \label{DsistributedMap}
\end{equation}
It is important to note that the resulting MAP problem has lower dimensions than the one handled without clusters. In spite of this, a major drawback comes from the fact that the matrix $\mathbf{S}_g$ from \eqref{zOpt} is of very low rank, and thus its pseudo-inverse results in a mismatched measurement, i.e.,$\hat{\mathbf{Y}}_g\neq \mathbf{Y}_g$, even under noise-free conditions. Motivated by the AEM method from inverse problems and Bayesian AEM in \cite{lunz2021learned}, we account for this mismatch by imposing data-driven model corrections. \textcolor{black}{Note that this data-driven correction is based on statistical learning and aims to estimate the mean and covariance of the mismatch error, different from machine learning techniques.} The following subsection provides a brief background of the AEM-inspired solutions.  
\subsection{Review of AEM}
This subsection briefly introduces the AEM \cite{lunz2021learned,arridge2006approximation,kaipio2006statistical}, which is used to develop the proposed JADCE algorithms. AEM is utilized to address the numerical error between $\hat{\mathbf{Y}}_g$ and $\mathbf{Y}_g$ resulting from the pseudo-inverse operation \eqref{pseudoInv}, particularly when $\mathbf{S}_g$ is low-rank, as considered in the sequel. For this reason, $\mathbf{\hat{Y}}_g$ is not a reliable measurement of the received signal corresponding to the $g$-th cluster. However, the aim of JADCE is to recover/estimate a hidden variable $\mathbf{X}_g$ using some set of measurements  while relying on their ideal linear relationship. This linear relationship is defined by a measurement matrix $\mathbf{S}_g$, which is assumed to be well known in advance and contains all the necessary information about how $\mathbf{X}_g$ is mapped into $\mathbf{Y}_g$. Such a relationship is modeled in noiseless scenarios by
\begin{equation}
\mathbf{Y}_g\!=\!\mathbf{S}_g\mathbf{X}_g. 
\label{measureLinear}
\end{equation}
%\textcolor{red}{It is assumed that the matrix $\bm{A}$ is known in advance and accurately represents the desired properties of the measuring device. This has a close relationship with the signal model \eqref{mtc}, i.e., the pilot sequences of the MTDs also serve as a measurement matrix, ensuring the precise recovery of $\bm{X}$ from the received signal.}
 To that end, relying solely on $\hat{\mathbf{Y}}_g$ to recover the hidden variable $\mathbf{X}_g$ is tantamount to using an incorrect linear model, thus, equivalent to using an incorrect measurement matrix.\par Let $\mathbf{\Tilde{S}}_g$ be the unknown and incorrect measurement matrix, then the relationship that yields $\hat{\mathbf{Y}}_g$ can be modeled by
\begin{equation}
\hat{\mathbf{Y}}_g\!=\!\mathbf{\Tilde{S}}_g\mathbf{X}_g,
\end{equation} 
and introduces a systematic model error 
 \begin{equation}
    \Delta \mathbf{Y}_g\!=\!\mathbf{Y}_g\!-\!\hat{\mathbf{Y}}_g.
    \label{errorModel}
\end{equation}
Failure to account for this error has a negative impact on the JADCE performance. Fortunately, this discrepancy can be corrected by leveraging AEM concepts from the field of inverse problems\cite{arridge2006approximation,lunz2021learned}. This is achieved by solving for $\mathbf{X}_g$ using a corrected version of the measurement matrix $\mathbf{\Tilde{S}}_g$, and thus providing a general framework for CS-MUD that can be applied in cases where the sensing matrix has errors \cite{wiesel2008linear}. Essentially, we can approximate $\mathbf{S}_g$  by ${\mathbf{C}_{g}\mathbf{\Tilde{S}}_g\approx \mathbf{S}_g}$, where $\mathbf{C}_g$ represents the correction term associated with the model error \eqref{errorModel} and that can be obtained through statistical training. Since the MTDs are considered to be stationary, it is possible to acquire each $\mathbf{C}_g$ offline. In addition, this training process does not contribute to the computational complexity of the JADCE solutions. Precisely,  $\mathbf{C}_g$ is computed from the covariance matrix of the model error, i.e., $\bm{\Phi}_g$, as will be discussed at the beginning of Section~\ref{AEMADMMSEC}. First, the model error can be acquired in each cluster  and considering $\tau$ training samples using
\begin{equation}    {\mathbf{E}_g(i)\!=\!\mathbf{S}_{g}\mathbf{X}_{g}(i)\!-\!\mathbf{\hat{Y}}_g}(i), i =1, \cdots, \tau.
    \label{Train}
\end{equation}
Here, $\mathbf{E}_g(i)= [\mathbf{e}_1(i),\cdots,\mathbf{e}_M(i)]$, where, ${\mathbf{e}_m(i) = \mathbf{S}_{g}\mathbf{x}_{m}(i) - \mathbf{\Tilde{S}}_g\mathbf{x}_m(i)}$, corresponds to the mismatch error at the $m$-th antenna for the $i$-th training sample. \textcolor{black}{By training} with \eqref{Train}, the average mismatch error for the $m$-th antenna is given by $\bm{\mu}_{m} =\sum_{i=1}^\tau \mathbf{e}_m (i)/\tau$. \textcolor{black}{Then,} the sample error covariance matrix of the mismatch error at the $m$-th antenna in a cluster is given by
\begin{equation}
    \bm{\Omega}_{m} = \frac{1}{\tau-1}\sum_{i=1}^{\tau}  \mathbf{e}_m(i)\mathbf{e}_{m}(i)\herm - \bm{\mu}_{m}\bm{\mu}_{m}\herm,
\end{equation} 
which converges to the population \textcolor{black}{error} covariance matrix as $\tau\rightarrow\infty$.  
\begin{remark}
    Given that the MTDs are stationary and by relying on the law of large numbers, the average covariance of the mismatch error for the $g$-th cluster is $\bm{\Phi}_g = \mathbb{E}\{\bm{\Omega}_{m}\}$. 
\end{remark}
Following this remark, we next introduce $\mathbf{C}_g$ into the optimization problem \eqref{inrodOptimization} and subsequently present the AEM-inspired JADCE solutions that are data-driven for each cluster. 
  % or the AEM based SSR that relies on $\bm{\Phi}_g$ to correct the mismatched measurement ($\hat{\bm{Y}_g}$) that is used at each cluster. devise data-driven solutions.f
%\textcolor{red}{

%\begin{equation}
  % \bm{Y}_i =\bm{A}_{i}\bm{X}_{i} +\bm{W}_{i}
  %  \label{equationClusterrs}
%\end{equation}
%The equation~\ref{equationClusterrs} can be vectored to formulate Bayesian recovery
%To facilitate this, we define the posterior distribution which results for the product of the two distributions as 

%\begin{equation}
    %p(\bm{X}_{i}|\bm{Y}) = %\frac{p(\bm{X}_{i})}{Z}\prod_{t=1}^{T}\frac{\mathrm{exp}\left(-(\bm{y}_{t}-\bm{a}\tran\bm{X})\Delta^{-1}(\bm{y}_{t}-\bm{a}\tran\bm{X})\herm\right)}{\pi^M|\Delta|},
%\end{equation}
%where the parameter $Z$ is the normalization constant\footnote{This is also defined as a partition function in convex optimization terminology} defined by,
%\begin{equation}
 %  Z = \prod_{t=1}^{T}p(\bm{X}\frac{\mathrm{exp}\left(-(\bm{y}_{t}-\bm{a}\tran\bm{X})\Delta^{-1}(\bm{y}_{t}-\bm{a}\tran\bm{X})\herm\right)}{\pi^M|\Delta|}d\bm{X}.
%\end{equation}}
%%%%%%%%%%%%%%%%%%

\subsection{Solution via AEM-ADMM}
\label{AEMADMMSEC}
The first AEM-inspired solution is derived via the ADMM framework that relies on the iterative soft threshold algorithm (ISTA). From the discussion of the AEM, we note that the noisy measurement for each cluster is related to the results of its pseudo-inverse by \begin{align}    \mathbf{Y}_{g}=\mathbf{S}_g\mathbf{X}_g+\mathbf{E}_g+\mathbf{W}      =\hat{\mathbf{Y}}_{g}+\mathbf{E}_g~,\forall g,
    \label{separatedMeasure} 
\end{align}
where $\mathbf{E}_g$ is the model error from \eqref{Train} and $\mathbf{W}$ comes from \eqref{mtc}. Given this, the net error from noise and the model error is defined as
$\bm{\Xi}_{g} = [\bm{\xi}_{1},\cdots,\bm{\xi}_{M}]$, such that $\bm{\Xi}_g = \mathbf{E}_g\!+\!\mathbf{W}$. By relying on the law of large numbers, we approximate each column of $\bm{\Xi}_{g}$ by a Gaussian variable, i.e., \textcolor{black}{$\bm{\xi}_{m} =\mathbf{e}_{1}\!+\!\mathbf{w}_{i}\sim \mathcal{CN}(\bm{\psi}_m,\bm{\Omega}_{m})$}, where $\bm{\psi}_m$ is acquired through training, similar to $\bm{\mu}_m$. Given its Gaussian nature, $\bm{\xi}_{m}$ has a precision matrix with the Cholesky decomposition $\mathbf{C}_g\tran\mathbf{C}_g\!=\!\bm{\Phi}_{g}^{-1}$ and thus facilitates the formulation of the exponential likelihood function\cite{lunz2021learned}  \begin{equation}
    \mathcal{P}(\mathbf{Y}_{g}|\mathbf{X}_{g})\propto \mathrm{exp}{\left( -\frac{1}{2}\norm{\mathbf{C}_{g}\left(\mathbf{S}_g\mathbf{X}_g\!-\!\hat{\mathbf{Y}}_{g}\!+\!\bm{\Psi}_{g}\right)}_{F}^{2} \right)}, 
    \label{likelihoodLearnt}
\end{equation}
where $\bm{\Psi}_g = [\bm{\psi}_1,\cdots,\bm{\psi}_M]$. \textcolor{black}{Hence}, the JADCE solution is  
\begin{align}
   \hat{\mathbf{X}}_g &\!=\!\operatorname*{argmax}_{\mathbf{X}_g} \mathcal{P}(\mathbf{Y}_g|\mathbf{X}_g)\mathcal{P}(\mathbf{X}_g)
   \nonumber \\ 
%&
&\!\overset{(a)}
=\!\operatorname*{argmax}_{\mathbf{X}_g}     \text{ln}\mathcal{P}(\mathbf{Y}_g|\mathbf{X}_g)+ \text{ln}\mathcal{P}(\mathbf{X}_g) 
\nonumber\\
 &\!\overset{(b)}=\!\operatorname*{argmin}_{\mathbf{X}_g}\frac{1}{2}\!\norm{\mathbf{C}_{g}\!\left(\mathbf{S}_g\mathbf{X}_{g}\! \!-\! \hat{\mathbf{Y}}_{g} \!\!+\! \bm{\Psi}_g\right)}_{F}^{2} \!\!+\! \lambda \!\norm{\mathbf{X}_g}_{2,1}\!,
 \label{admmProblem}
\end{align}
where (a) comes from using the logarithmic form, while (b) leverages \eqref{likelihoodLearnt} while taking $\lambda\lVert \mathbf{X}_g\rVert_{2,1}$ as an approximation of $-\text{ln}\mathcal{P}(\mathbf{X}_g)$. Notice that $\lambda$ trades off between the measurement fidelity and the sparsity of the solution. 

To implement \eqref{admmProblem} via the ADMM framework \cite{djelouat2021spatial,boyd2011distributed}, we \textcolor{black}{reformulate \eqref{admmProblem}} for each cluster $g \in \mathcal{G}$ as follows
\begin{subequations}\label{adMMfirst} 
\begin{align}
       \operatorname*{minimize}_{\mathbf{X}_g, \mathbf{Z}_g}& \quad  \frac{1}{2}\norm{\mathbf{C}_g\left(
 \mathbf{S}_{g}\mathbf{Z}_g-\hat{\bm{Y}}_{g} + \bm{\Psi}_{g}\ \right)}_{F}^2 \nonumber \\ &\, \hspace{1.5 cm} + \lambda \norm{\mathbf{X}_g}_{2,1} + \frac{\rho}{2}\norm{\mathbf{X}_g-\mathbf{Z}_g}_F^2\label{x_0} \\\
      \text{subject to} & \quad  \mathbf{X}_g = \mathbf{Z}_g \label{Rank},
  \end{align}
\end{subequations}
where $\rho$ is the ADMM 
 step size and $\mathbf{Z}_g = [\mathbf{z}_{n,g},\cdots,\mathbf{z}_{N_g,g}]\tran$ is the auxiliary variable that facilitates the closed form update of the estimate of $\mathbf{X}_g$ through the Moreau-Yosida regularization \cite{boyd2011distributed}. Following the formulation \eqref{adMMfirst}, the augmented Lagrangian is expressed as \cite{boyd2011distributed}
\begin{align} \mathcal{L}\left(\mathbf{X}_g,\bm{\Theta}_{g}\right)&\overset{(a)}{=}\frac{1}{2}\norm{\mathbf{C}_g\left(
 \mathbf{S}_{g}\mathbf{Z}_g\!-\!\hat{\mathbf{Y}}_{g} \!+\!\bm{\Psi}_{g}\right)}_{F}^2\!+\! \lambda \norm{\mathbf{X}_g}_{2,1}\!\nonumber\\
 &\, \hspace{1.5cm}+\! \bm{\Theta}_{g}\tran\left(\mathbf{X}_g\!-\!\mathbf{Z}_g\right) + \frac{\rho}{2}\norm{\mathbf{X}_g-\mathbf{Z}_g}_{F}^{2}\nonumber\\
&\overset{(b)}{=}  \frac{1}{2}\norm{\mathbf{C}_g\left(
 \mathbf{S}_{g}\mathbf{Z}_g\!-\!\hat{\mathbf{Y}}_{g} \!+\!\bm{\Psi}_{g}\right)}_{F}^2\!+\! \lambda\norm{\mathbf{X}_g}_{2,1}\!\nonumber\\
 &\hspace{1.5cm}+\!\norm{\mathbf{X}_g\!-\!\mathbf{Z}_g\!+\!\frac{\bm{\Theta}_{g}}{\rho}}_{F}^{2}\!-\!\frac{\norm{\bm{\Theta}_{g}}_{F}^{2}}{2\rho}, 
\end{align}  
where $\bm{\Theta}_{g} = [\bm{\theta}_{1,g},\cdots,\bm{\theta}_{N_g,g}] \in \mathbb{C}^{M \times N_g}$ is the dual matrix for the augmented Lagrangian. The expression of $\mathcal{L}\left(\mathbf{X}_g,\bm{\Theta}_g\right)$ in $(a)$ is the standard augmented Lagrangian function, while $(b)$ is the scaled Lagrangian form \cite{boyd2011distributed}. The JADCE is solved via AEM-ADMM by updating the set of variables $\{\mathbf{X}_g$,$\mathbf{Z}_g$,$\bm{\Theta}_g\}$ in an alternating manner. Precisely, the sub-problems corresponding to the variables are each given in the ${(t+1)}$-th iteration by 
\begin{algorithm}[t]
  \KwIn{$\mathbf{\hat{Y}}_g$, $\Delta$, $\mathbf{C}_g$, $\bm{\Psi}_g$}
   
  Initialisation: $\mathbf{X}_{g}^{(0)} = \bm{0}$, $\mathbf{Z}_g^{(0)}=\bm{0}$, $\bm{\Theta}_g^{(0)}=\bm{0}$, $t= 0$\\\
    \Repeat{$\frac{\norm{\mathbf{Z}_{g}^{(t+1)}-\mathbf{Z}_{g}^{(t)}}_{F}}{\norm{\mathbf{Z}_{g}^{(t)}}_{F}} < \Delta$}{%
      \label{stepA}
      
   Update $\mathbf{Z}_{g}^{(t+1)}$ using \eqref{zDerive}\
   
    Update $\mathbf{X}_{g}^{(t+1)}$ using \eqref{xHaT}\ 
    
    $\bm{\Theta}_g^{(t+1)} = \bm{\Theta}_{g}^{(t)} + \rho \left(\mathbf{X}_{g}^{(t+1)} - \mathbf{Z}_{g}^{(t+1)} \right)$\   
    
       $t = t + 1$\
    }
    \KwOut{$\hat{\mathbf{X}}_{g} = \mathbf{Z}_g$ }
  \caption{AEM-ADMM,   $\forall g \in \mathcal{G}$}
  \label{algoADMM}
\end{algorithm}
\begin{align} \mathbf{Z}^{(t+1)}_{g}&=\!\min_{\mathbf{Z}_g}\mathcal{L}\left(\mathbf{X}_g^{(t)},\mathbf{Z}_g,\bm{\Theta}_{g}^{(t)}\right) \nonumber \\ 
& \!=\!\min_{\mathbf{Z}_g}\frac{1}{2}\norm{\mathbf{C}_g\left(
 \mathbf{S}_{g}\mathbf{Z}_g\!-\!\hat{\mathbf{Y}}_{g}\!+\! \bm{\Psi}_g\right)}_{F}^2  \nonumber\\ 
 &\, \hspace{3cm}  \!+\!\norm{\mathbf{X}_g^{(t)}\!-\!\mathbf{Z}_g\!+\!\frac{\bm{\Theta}_{g}^{(t)}}{\rho}}_{F}^{2}\!, 
\label{zUpdate}\\
\mathbf{X}^{(t+1)}_{g}&=\min_{\mathbf{X}_g}\mathcal{L}\left(\mathbf{X}_g,\mathbf{Z}_g^{(t+1)},\bm{\Theta}_{g}^{(t)}\right) \nonumber\\ 
& =\min_{\mathbf{X}_g}  \norm{\mathbf{X}_g}_{2,1}+\frac{\rho}{2} \|\mathbf{X}_g-\mathbf{Z}_g^{(t+1)}+ \frac{1}{\rho}\bm{\Theta}_{g}^{(t)} \|_{\mathrm{F}}^2,
\label{xUpdate}\\
\bm{\Theta}_g^{(t+1)} &= \bm{\Theta}_g^{(t)}+\rho\big(\mathbf{X}_g^{(t+1)}  -\mathbf{Z}_g^{(t+1)} \big),
\label{lamdaUpdate}
\end{align}
such that $\mathbf{Z}^{(t+1)}_{g}$ is updated by minimizing \eqref{zUpdate} with respect to $\mathbf{Z}_g$ while holding all the other variables constant. That is, computing the derivative with respect to $\mathbf{Z}_g$, setting it to zero and solving for $\mathbf{Z}_g$ as follows
\begin{align}
   \frac{\partial \mathcal{L}\left(\mathbf{X}_g^{(t)},\mathbf{Z}_g,\bm{\Theta}_g^{(t)}\right) }{\partial \mathbf{Z}_g}& \nonumber\\
   =  \mathbf{S}_g\herm&\mathbf{C}_g\herm\mathbf{C}_g\mathbf{Z}_{g}+ \rho\mathbf{Z}- \rho\mathbf{X}_g^{(t)}+\bm{\Theta}_g^{(t)}  \nonumber\\  
 - &\mathbf{S}_g\herm\mathbf{C}_g\herm\mathbf{C}_g\mathbf{\hat{Y}}_{g}+\mathbf{S}_g\herm\mathbf{C}_g\herm\mathbf{C}_g\bm{\Psi}_g = \mathbf{0},
   \label{derivativeX}
\end{align}
which leads to 
\begin{align}
\mathbf{Z}_g^{(t+1)} =& \left(\mathbf{S}_g\herm\mathbf{C}_g\herm\mathbf{C}_g\mathbf{S}_g + \rho\mathbf{I}_{N_g}\right)^{-1}\times\nonumber\\
&\left(\rho\mathbf{X}^{(t)}\!\!-\! \mathbf{\Theta}_g^{(t)}\!+\!\mathbf{S}_g\herm\mathbf{C}_g\herm\mathbf{C}_g\mathbf{\hat{Y}}_{g}\!\!-\!\mathbf{S}_g\herm\mathbf{C}_g\herm\mathbf{C}_g\bm{\Psi}_g\right)\!.
\label{zDerive}
\end{align}
Similarly, the computation of $\mathbf{X}_g^{(t+1)}$ involves minimizing \eqref{xUpdate} with respect to $\mathbf{X}_g$. By observing that \eqref{xUpdate} is the Moreau envelope of the mixed norm $\norm{\mathbf{X}_{g}}_{2,1}$, we can update $\mathbf{X}_g$ based on the results obtained from \eqref{zDerive}. Therefore, in order to enforce the sparsity of the solution to JADCE, the $\ell_2$ norms of the rows of $\mathbf{X}_g$ must be sparse. We can denote the norms of the rows of $\mathbf{X}_g$ by $\mathbf{\bar{x}}_g\in \mathbb{R}^{N_g\times 1}$, \textcolor{black}{such that we} update $\mathbf{X}_g$  by solving 
\begin{equation}
\mathbf{X}_{g}^{(t+1)}=\min_{\mathbf{X}_g} \norm{\mathbf{\bar{x}}_g}_{1}+\frac{\rho}{2} \norm{\mathbf{X}_{g}- \bm{\Pi}_{g}^{(t+1)}}_{F}^2, 
    \label{lassoMinimize}
\end{equation}
where $\bm{\Pi}_{g}^{(t+1)} = \mathbf{Z}_{g}^{(t+1)}+ \frac{1}{\rho}\bm{\Theta}_{g}^{(t)}$, and using the proximal operator method \cite{boyd2011distributed}, one obtains $N_g$ decoupled solutions \cite{goldstein2014field} 
\begin{align}
\mathbf{x}_{n,g}^{(t+1)} &= \text{prox}_{\lambda,\norm{\mathbf{\bar{x}}_g}_{1}}(\bm{\Pi}_g^{(t+1)},\rho)  \nonumber\\ 
& =\! \bm{\pi}_{n,g}^{(t+1)}\frac{\text{max}\{\lVert\bm{\pi}_{n,g}^{(t+1)}\!\rVert\!-\! \rho^{-1}\!, 0\}}{\lVert\bm{\pi}_{n,g}^{(t+1)}\rVert},  n \!=\! 1,\!\cdots\!, N_g.
 \label{xHaT}   
\end{align}
% \begin{equation}    \mathbf{x}_{n,g}^{(t+1)}= \text{prox}_{\lambda,\norm{\mathbf{\bar{x}}_g}_{1}}(\bm{\pi}_{n,g},\lambda) = \begin{cases} 
%     \bm{\pi}_{n,g}-\bm{\lambda},~\text{if $\bm{\pi}_{n,g}>\lambda$} \\
%        \bm{\pi}_{n,g}+\bm{\lambda},~\text{if $\bm{\pi}_{n,g}<\lambda$} \\0,~\text{otherwise}\\
%    \end{cases}. 
%    \label{xHaT}
% \end{equation}
A summary of AEM-ADMM is presented in Algorithm~\ref{algoADMM}. Note that the AEM-ADMM inherits the properties of classical ADMM \cite{boyd2011distributed} and thus has slow convergence to the best possible accuracy, even though, most of its results are acceptable for the JADCE framework. Next, we present the AEM-SBL which is the Bayesian solution and thus exploits the statistical distributions of the observations and the prior.  
\subsection{Solution via AEM-SBL}
\label{AEMsblSEC}The AEM-ADMM developed in Section \ref{AEMADMMSEC} relies on a sparsity promoting penalty and not on the explicit prior distribution $\mathcal{P}(\mathbf{X}_g)$, which may be highly sub-optimal. Therefore, we introduce the AEM-SBL, another AEM-inspired JADCE solution that relies on the SBL framework \cite{zhang2011sparse,rajoriya2022centralized}. Given its Bayesian nature, AEM-SBL exploits $\mathcal{P}(\mathbf{X}_g)$, which is a clear advantage over AEM-ADMM. The AEM-SBL uses the joint posterior distribution \begin{align}
     \mathcal{P}(\mathbf{X}_{g},\!\bm{\Gamma}_{g}|\mathbf{Y}_{g}) \!&=\!\frac{\mathcal{P}(\mathbf{Y}_{g}|\mathbf{X}_{g},\!\bm\Gamma_{g})\mathcal{P}(\mathbf{X}_{g}|\bm{\Gamma}_{g})\mathcal{P}(\bm{\Gamma}_{g})}{\mathcal{P}(\mathbf{Y}_{g})} \nonumber\\     
 &
 \!=\! \frac{\mathcal{P}(\mathbf{\hat{Y}}_{g}\!+\! \mathbf{\Psi}_g|\mathbf{X}_{g},\!\bm\Gamma_{g})\mathcal{P}(\mathbf{X}_{g}|\bm{\Gamma}_{g})\mathcal{P}(\bm{\Gamma}_{g})}{\mathcal{P}(\mathbf{Y}_{g})}, %\forall g,
 \label{centBayes}
\end{align}  
  where the second \textcolor{black}{step} follows from the corrected likelihood \eqref{likelihoodLearnt}, for which $\mathcal{P}(\bm{y}_m|\mathbf{x}_m)\approx \mathcal{CN}(\mathbf{S}_m\mathbf{x}_m,\bm{\Phi}_g)$ as a result of AEM, while $\bm{\Gamma}_g = \text{diag}\{\bm{\gamma}_{g}\}$, where $\bm{\gamma}_{g} = \{\gamma_{1,g},\cdots,\gamma_{N_g,g}\}\in \mathbb{R}_{+}^{N_g\times 1}$ are the sparsity promoting hyper-parameters in each cluster. Notice that the diagonal precision matrix $\mathbf{\Gamma}_g$ of these hyper-parameters creates a fixed sparsity pattern across each row of $\hat{\mathbf{X}}_{g}$, and thus allowing the decomposition of the joint posterior distribution in each cluster \textcolor{black}{as}
\begin{align}  
\mathcal{P}(\mathbf{X}_{g},\bm{\Gamma}_{g}|\mathbf{Y}_{g}) \propto & \prod_{m=1}^{M}\mathcal{P}\left(\hat{\mathbf{y}}_{m}\!+\!\bm{\psi}_m|\mathbf{S}_{g}\mathbf{x}_{m}  \right)\times \nonumber\\
&\prod_{n=1}^{N_g}\mathcal{P}\left(\mathbf{x}_{n,g}|\gamma_{n,g} \right)\!%\times\!
\prod_{n=1}^{N_g}\mathcal{P}\left(\gamma_{n,g} \right).
    \label{bethe}
 \end{align}
\textcolor{black}{This enables the} independent update of  $\gamma_{n,g}$ and $\mathbf{x}_{n,g}$ as will be shown later. \par Following the conventional SBL framework, the AEM-SBL uses \eqref{bethe} to find $\bm{\gamma}_{g}$ and $\mathbf{X}_{g}$ using AEM corrected expectation (E) step and the maximization (M)-step, respectively \cite{zhang2011sparse,sant2022block}. To facilitate this, we define the AEM-SBL cost as a function of $\bm{\gamma}_{g}$ at each cluster, i.e., 
 \begin{align}
\bm{\gamma}_{g} &=     \underset{\bm{\gamma}_g}{\mathrm{argmax}}~\mathrm{ln}\,\mathcal{P}( \bm{\gamma}_g|\mathbf{\hat{Y}}_{g}+ \mathbf{\Psi}_g)
\nonumber \\
 & \propto   \underset{\bm{\gamma}_g}{\mathrm{argmax}}~\mathrm{ln}\,\mathcal{P}(\mathbf{Y}_{g}|\bm{\gamma}_g)\mathcal{P}(\bm{\gamma}_g).   
   \label{sblCost} 
\end{align}
Note that the problem \eqref{sblCost} requires the marginalization of $\mathcal{P}(\mathbf{X}_{g},\mathbf{Y}_{g},\bm{\gamma}_{g})$ with respect to $\mathbf{X}_{g}$, which can be solved iteratively using EM. Therefore, in the $(t+1)$-th iteration, the corrected E-step is computed using the log-likelihood of the complete joint distribution with respect to the posterior distribution that is parameterized on the previous estimate of $\bm{\gamma}_g$, i.e., $\mathcal{P}(\mathbf{x}_m|\mathbf{y}_m,\bm{\gamma}_g^{(t)})$. Let this expectation be defined  by \begin{align}   J(\bm{\gamma}_g,\bm{\hat{\gamma}}_g^{(t)}) &=  \mathbb{E}_{\mathcal{P}(\mathbf{x}_m|\mathbf{y}_m,\bm{\hat{\gamma}}_g^{(t)})}\text{ln}\mathcal{P}(\mathbf{x}_m,\mathbf{y}_m|\bm{\gamma}_g)\nonumber\\
&=  \mathbb{E}_{\mathcal{P}(\mathbf{x}_m|\mathbf{y}_m,\bm{\gamma}_g^{(t)})}[\text{ln}\mathcal{P}(\mathbf{x}_m|\bm{\gamma}_g)  \nonumber\\ 
&\,\hspace{0.5cm}  + \text{ln}\mathcal{P}(\mathbf{y}_m|\mathbf{x}_m) + \text{ln}\mathcal{P}(\bm{\gamma}_g)], \forall m \in \mathcal{M}.
\label{sblProcedure}
\end{align} 
The decomposition presented in \eqref{bethe} allows expressing the complete expectation as follows
\begin{equation}   J(\bm{\gamma}_g,\bm{\hat{\gamma}}_g) \propto  \sum_{n=1}^{N_g}\text{ln}\gamma_{n,g} - \mathbb{E}_{\mathcal{P}(\mathbf{x}_g|\mathbf{y}_g,\bm{\gamma}_g^{(t)})}[\norm{\mathbf{x}_{n,g}}^2],
\label{sblSimnple}
\end{equation} 
where the posterior distribution $\mathcal{P}(\mathbf{x}_m|\mathbf{y}_m,\bm{\hat{\gamma}}_g^{(t)})$ of \eqref{sblProcedure} is parameterized in the AEM framework by the corrected mean and covariance, respectively given by 
\begin{align}
     \mathbf{\hat{x}}_m&= \mathbf{\Sigma}_g\mathbf{S}_{g}\herm\mathbf{\Phi}_{g}^{-1}\left(\hat{\mathbf{y}}_m +\bm{\psi}_m \right), \forall m \in \mathcal{M}, \nonumber 
     \\
    \mathbf{\Sigma}_g&= \left(\mathbf{S}_{g}\herm\mathbf{\Phi}_{g}^{-1}\mathbf{S}_g\!+\!\bm{\Gamma}_g^{(t)}\right)^{-1}. 
    \label{EqautionESBL}
\end{align}
  Note that each diagonal entry of $\mathbf{\Sigma}_g$, which we denote by $\{\nu_{n,g}\}$ is common for all the entries of $\mathbf{x}_{n,g}$. Further, the second term of \eqref{sblSimnple} can be simplified by \cite{al2017gamp}
 \begin{align}
     \mathbb{E}_{\mathcal{P}(\mathbf{x}_g|\mathbf{y}_g,\bm{\hat{\gamma}}_g^{(t)})}[\norm{\mathbf{x}_{n,g}}^2] &= \mathbb{E}_{\mathcal{P}(\mathbf{x}_g|\mathbf{y}_g,\bm{\gamma}_g^{(t)})}[\norm{\mathbf{x}_{n,g}}]^2 \nonumber\\ 
&\,\hspace{1cm}+ \mathbb{V}_{\mathcal{P}(\mathbf{x}_g|\mathbf{y}_g,\bm{\gamma}_g^{(t)})}[\mathbf{x}_{n,g}],
 \end{align} 
 thus, for the $n$-th device in the $g$-th cluster  
 \begin{equation}   J(\gamma_{n,g},\hat{\gamma}_{n,g})\propto  \text{ln}\gamma_{n,g} -\gamma_{n,g}(\norm{\mathbf{\hat{x}}_{n,g}}^{2} + \nu_{n,g}).
\label{simpleSbl}
\end{equation}
 In the M-step, the hyper-parameters are computed for each MTD by solving \eqref{sblCost} as follows
 \begin{equation}
   \frac{\partial J\left(\gamma_{n,g},\gamma_{n,g}^{(t)}\right) }{\partial\gamma_{n,g}} = \frac{1}{\gamma_{n,g}} - (\norm{\mathbf{\hat{x}}_{n,g}}^2 + \nu_{n,g}) = 0,   
 \end{equation}
 which leads to \begin{equation}
    \gamma_{n,g}^{(t+1)} = \frac{1}{\norm{\mathbf{\hat{x}}_{n,g}}^2 + \nu_{n,g}},\qquad \forall g,n. \label{gammaUpdate} \end{equation}  
   A summary of the AEM-SBL is given by Algorithm \ref{algoAEMSBL}.
\begin{algorithm}[!t]
  \KwIn{$\mathbf{\hat{Y}}_g$, $\Delta$, $\mathbf{C}_g$}
 % \KwOut{$\hat{\bm{X}}^t$}
  Initialization: $\mathbf{X}^0 = \bm{0}$, 
  $\mathbf{X}^0 = \bm{0}$, $\bm{\Gamma}^{(0)} = \bm{1}_{N_g\times 1}$, $t= 0$\\\
    \Repeat{$   \frac{\norm{\mathbf{X}_{g}^{(t+1)}-\mathbf{X}_{g}^{(t)}}_{F}}{\norm{\mathbf{X}_{g}^{(t+1)}}_{F}} < \Delta$}{%
      %\label{stepB}
      Update $\mathbf{X}_{g}^{(t+1)}$ using \eqref{EqautionESBL} \
     
   Update $\bm{\Gamma}^{(t+1)}$ using \eqref{gammaUpdate} \
     
       $t = t + 1$\
       
    }
     \KwOut{$\mathbf{\hat{X}_{g}} = \mathbf{X}_{g}^{(t)}$}
  \caption{AEM-SBL, $\forall g \in \mathcal{G}$}
  \label{algoAEMSBL}
\end{algorithm}
\begin{table}[t!]
\centering
\caption{Computational complexity of  JADCE algorithms}
\small
\begin{tabular}{c||c||c} 
 \hline
  \textbf{Algorithm} & \textbf{No. operations per iteration} &  \textbf{Complexity} \\ [0.5ex] 
 \hline\hline
ADMM & $N^2M + NM^2 + MLN$&$\mathcal{O}(N^2M \!+\! NM^2)$\\AEM-ADMM & $\frac{N^2M}{G^2} + \frac{NM^2}{G} + MLN$ & $\mathcal{O}(\frac{N^2M}{G^2} \!+\! \frac{NM^2}{G})$\\  
SBL & $N^2L + N^2$ + $
NM$& $\mathcal{O}(N^2L)$ \\
AEM-SBL & $\frac{N^2L}{G} + \frac{N^2}{G}$ + $
NM$ & $\mathcal{O}(\frac{N^2L}{G})$\\
SOMP  &  $(2L \!+\! 1)MN\! +\! L(M^2 \!+\! M \!+\!1)$& $\mathcal{O}(LMN)$  \\
 CB-SOMP & $(2L \!+\! 1)MN \!+\! L(M^2 \!+\! M \!+\!1)$ & $\mathcal{O}(LMN)$ \\ [1ex] 
 \hline
\end{tabular}
\label{complexAnalysis}
\end{table}
\subsection{Complexity Analysis} The computational complexity of the algorithms is given in  Table~\ref{complexAnalysis} in terms of the big-$\mathcal{O}$ notation, which considers relevant mathematical operations, e.g., matrix multiplications and inversions. \textcolor{black}{Motivated by the dependency of AEM-ADMM and AEM-SBL on the pre-processing steps \eqref{decorelated} and \eqref{pseudoInv}, we analyze their computational complexity, which is $\mathcal{O}(\frac{NLM}{G})$ and $\mathcal{O}\left(\frac{L^2N}{G} + L^3  + \frac{NLM}{G}\right)$, respectively. The most computationally expensive operation in the pre-processing is $(\mathbf{S}_{g}\mathbf{S}_{g}\herm)^{-1}\mathbf{S}_{g}$, which has a complexity of $\mathcal{O}\left(\frac{L^2N}{G} + L^3\right)$. Fortunately, this operation can be pre-computed and remain valid until the pilot allocation changes e.g., when the number of devices changes, consequently reducing the complexity of the pre-processing step to $\mathcal{O}(\frac{NLM}{G})$}. All in all, notice that \textcolor{black}{$\mathcal{O}(\frac{NLM}{G})$} goes down as the number of clusters increases, \textcolor{black}{and thus} AEM-ADMM and  AEM-SBL have reduced computational complexity compared with their conventional implementations. To this end, the proposed cluster-wise AEM algorithms are bound to have shorter runtime as will be seen in the \textcolor{black}{next section.} 
\section{Numerical results and discussions}
\label{results}
In this section, we present the results of the proposed AEM-ADMM and the AEM-SBL in comparison to other JADCE approaches. The numerical results are presented in terms of the channel estimation accuracy, detection capabilities, and scalability via the normalized mean squared error (NMSE), the average probability of miss detection (PMD), and the algorithm run-time, respectively.  To clarify, we will only present the performance metrics for cluster-wise performance as they can be easily extended to evaluate the performance of centralized algorithms such as ADMM, SOMP, and SBL.\label{performanceMetrics}
\par The channel estimation accuracy is evaluated using the NMSE defined by
\begin{equation}
  \text{NMSE}_{i}\!=\!\mathbb{E}_{i,g} \left(\frac{\norm{\mathbf{x}_{i,g} - \mathbf{\hat{x}}_{i,g}}^2}{\norm{\mathbf{x}_{i,g} }^2}\right),  i\in\mathcal{K}_g. 
  \label{channelMtds}
\end{equation}
For PMD, we first estimate the activity vector $\hat{\bm{\alpha}}_g\in \{0,1\}^{N_g\times 1}$ using
\begin{equation}
  \hat{\alpha}_{n,g}\!=\!
  \begin{cases} 
      1,~\text{if $\norm{\mathbf{\hat{x}}_{n,g}}_{2}\geq \zeta$}\\
      0,~\text{otherwise}\\
   \end{cases}, 
   \label{maximumThreshold}
\end{equation}
where $\zeta$ is a threshold that is set according to a fixed target probability of false alarm (PFA) (see Table~\ref{simulationPar}). It then follows that the PMD is computed using\begin{equation}
    \text{PMD}= \mathbb{E}\left( \frac{\sum_{n=1}^{N_{g}}\text{max}(0,\alpha_{n,g} - \hat{\alpha}_{n,g})}{|\mathcal{K}_{g}|} \right).
    \label{pmd}
\end{equation}
The run-time performance is evaluated in seconds, thus related to the number of iterations and the computational complexity analysis that was presented in Table~\ref{complexAnalysis}.  
\begin{table}
    \centering
  \caption{Simulation parameters}
 \begin{tabular}{  m{14em}  m{10em} } 
\hline
\textbf{Parameter} & \textbf{Value}\\ 
\hline
Cell radius & $250$ m \\ 
Number of MTDs $(N)$ & $1000$ \\
Number of clusters$(G)$ & $4$ \\
Bandwidth  & $20$ MHz  \\ 
Noise power ($\sigma^2$) & $2\!\times\!10^{-13}$ W \\ 
Coherence interval $(T)$ & $300$ \\ 
Length of the pilot sequences $(L)$ & $64$ \\ 
Number of BS antennas $(M)$ & $32$ \\ 
Activation probability $(\epsilon)$ & $0.01$ \\ 
$\text{Average SNR}$ & $10$ dB \\ 
Error tolerance $(\Delta)$ & $10^{-4}$ \\ 
Target $\text{PFA}$ & $10^{-3}$ \\ 
\hline
\end{tabular}
\label{simulationPar}
\end{table}
\begin{figure}[t!]
\centering
%\begin{subfigure}{0.5\textwidth} 
 \includegraphics[width=0.475\textwidth]{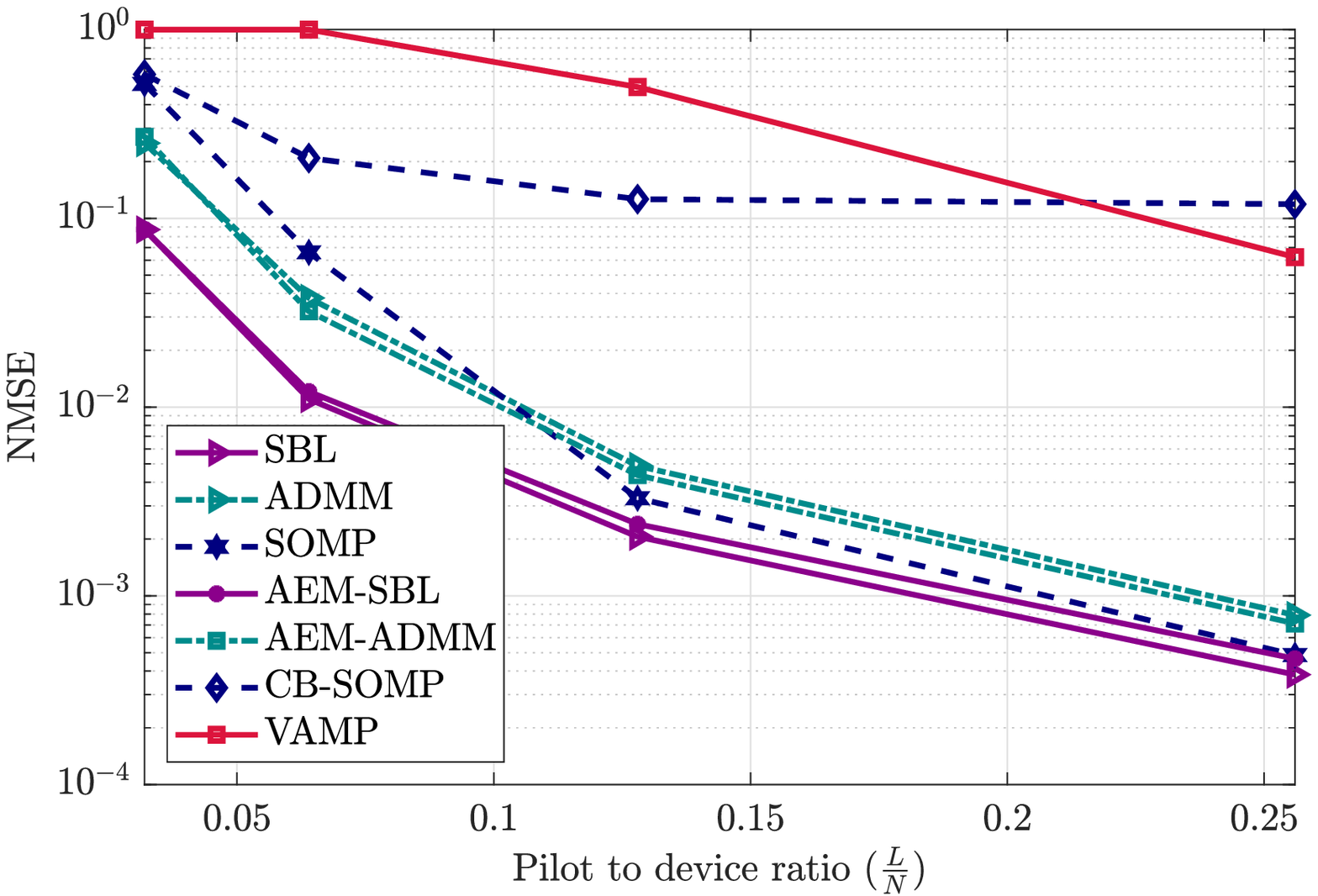}\\
 %\caption{~}
 %   \label{NMSEPILOT}
%\end{subfigure}
%\hfill
%\begin{subfigure}{0.5\textwidth}    
    \includegraphics[width=0.475\textwidth]{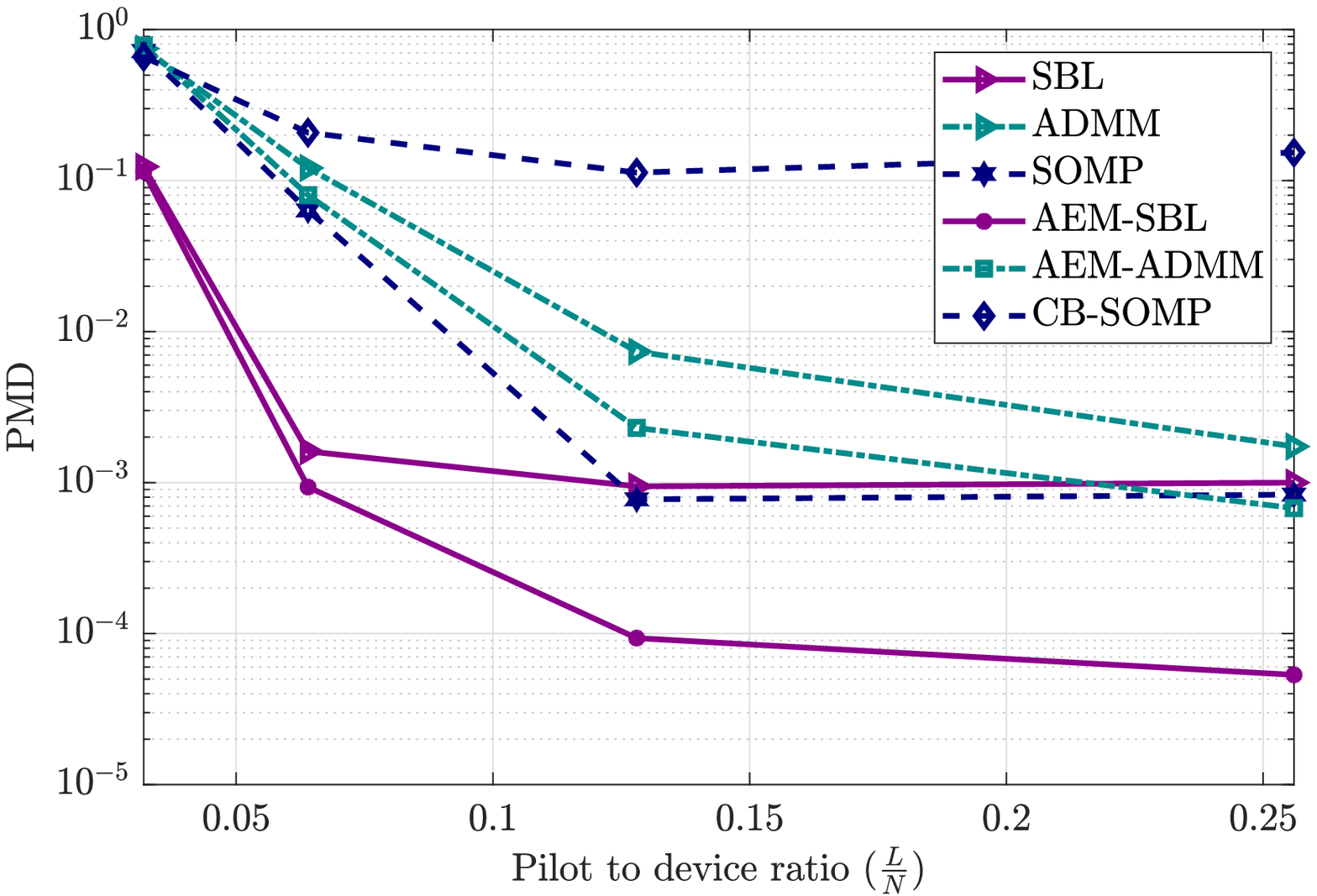}\\ 
%\caption{~}
 %\label{pmdToPilot}
%\end{subfigure}
%\hfill
%\begin{subfigure}{0.5\textwidth}   
    \includegraphics[width=0.475\textwidth]{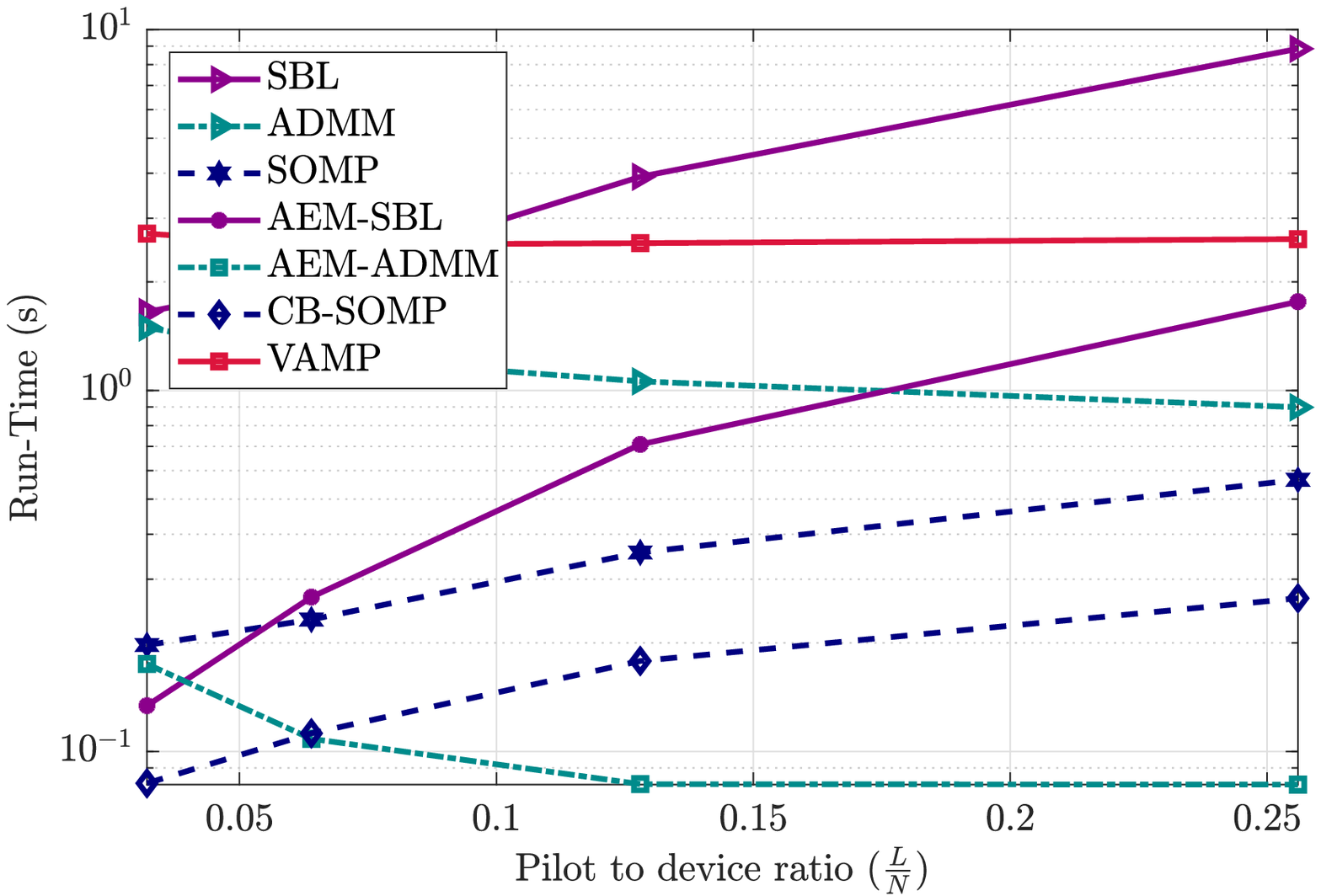}
%\caption{~}   
%\label{RuntimePilots}
%\end{subfigure}
\caption{Performance in terms of a) NMSE \textcolor{black}{(top), b) PMD (middle),  and c) run-time (bottom)} as a function of the pilot length.}
\label{pilotSequences}%\vspace{-7mm}
\end{figure}
\subsection{Simulation Setup}
 \textcolor{black}{We consider a single-cell massive MIMO uplink network with a radius of $250$ m, where the BS serves $1000$ MTDs randomly placed within the cell. We assume that the MTDs are uniformly partitioned into $G = 4$ clusters a prior. For simplicity, we adopt a log-distance path loss model such that $\beta_{n,g}=-130-37.6\log_{10} d_{n,g}$, where $d_{n,g}$ is the distance between the $n$-th MTD in 
 the $g$-th cluster and the BS. We consider two different setups: i) uncorrelated channels, for which $\mathbf{Q}_{n,g} = \beta_{n,g}\mathbf{I}$ and ii)  spatially correlated channels, for which the matrix ${\mathbf{Q}_{n,g}}$ is generated using the approximated Gaussian local scattering model with a half-wavelength antenna separation as described in \cite{bjornson2017massive}. The  $k$-th row and $m$-th column entries of the matrix are computed by 
\begin{align}
[\mathbf{Q}_{n,g}]_{k,m} \!=\! \frac{\beta_{n,g}}{L_p} \sum_{i=1}^{L_p} &\exp(\pi j(k\!-\!m)\sin\phi_{n,g}^{(i)})\times\nonumber\\
&\exp(-\frac{1}{2}\sigma_{\phi_{n,g}}^{2}\pi (k\!-\!m)\cos\phi_{n,g}^{(i)}),
\end{align}
where $L_p$ is the number of multi-path components. Additionally,  $\phi_{n,g}^{(i)} \sim \mathcal{U}(\Bar{\phi}_{n,g} - \frac{2\pi}{9}, \Bar{\phi}_{n,g} + \frac{2\pi}{9})$ is the nominal angle of the $i$-th multipath cluster distributed around the azimuth angle $\Bar{\phi}_{n,g}$ of the $n$-th MTD in the $g$-th cluster relative to the boresight of the BS antenna array. The angular standard deviation of the paths within the multipath cluster is denoted by $\sigma_{\phi_{n,g}}$, while complex Hadamard matrix is adopted as the basis matrix $\mathbf{B}$ of the pilot sequences.} Unless otherwise stated, the simulations are performed using the parameters provided in Table~\ref{simulationPar}. The figures display results obtained by averaging over $10^3$ Monte Carlo simulations.   
\subsection{On the length of the pilot sequences}
In Fig.~\ref{pilotSequences}, we assess the impact of the pilot lengths on the performance of the JADCE algorithms. Specifically, Fig.~\ref{pilotSequences}(a) illustrates how the NMSE varies with the under-sampling ratio, which measures the ratio of the pilot length to the number of devices in the network, i.e., $\frac{L}{N}$. In general, the results show that the channel estimation improves with the pilot length-to-device ratio. This is primarily because higher values of $\frac{L}{N}$ enable the BS to allocate more resources during the training phase, thereby improving the accuracy of the CSI. In addition, a higher $\frac{L}{N}$ ratio results in longer pilot sequences, which can promote orthogonality between the pilot sequences of different MTDs, leading to fewer pilot collisions in the network and a higher quality of the channel estimate. Accordingly, as $\frac{L}{N}$ increases, the JADCE algorithms benefit from improved RIP of the matrix $\mathbf{S}_g$. It is noteworthy that both AEM-SBL and AEM-ADMM exhibit similar performance to ADMM and SBL while solving smaller-sized problems. Therefore, both AEM-ADMM and AEM-SBL are highly efficient methods for performing channel estimation under practical conditions in mMTC, particularly when the pilot length is much smaller than the number of devices in the network, i.e., with $L\ll N$. From the same results, CB-SOMP (Algorithm~\ref{algoOMP}) exhibits poorer channel estimation capabilities than the classical SOMP. This underperformance of CB-SOMP can be attributed to its reliance on a completely mismatched model, thereby highlighting the importance of accounting for such mismatches, as demonstrated by our AEM-inspired algorithms (AEM-ADMM and AEM-SBL). Conversely, while SOMP and ADMM do not leverage the clustering structure, they exhibit strong performance under larger pilot lengths, which is impractical for mMTC applications. \textcolor{black}{From the results, it is evident that both the SBL and AEM-SBL algorithms significantly outperform VAMP. Although VAMP has high efficiency with near-Gaussian sensing matrices, it is limited in this setting due to the departure of the sensing matrix's structure from the Gaussian assumption \cite{senel2018grant}. Furthermore, as discussed in Section~\ref{literatureReview}, VAMP and other AMP-related algorithms are sub-optimal in problem dimensions such as the currently considered short pilot lengths, which are practical for mMTC.}

\par Fig.~\ref{pilotSequences}(b) shows how the PMD is affected by $\frac{L}{N}$. In general, the PMD decreases with the pilot-to-device ratio. In connection with the results of Fig.~\ref{pilotSequences}(a), this is due to the reduced pilot collisions of the different MTDs. Remarkably, the AEM-SBL outperforms the conventional SBL under this metric as it benefits from reduced inter-cluster MUI, making it efficient in imposing the row sparsity in the  recovery of the matrix $\mathbf{X}_g$. Furthermore, AEM-ADMM outperforms ADMM in terms of this metric by efficiently handling the MUI, thereby providing a low-complexity alternative JADCE solution for clustered mMTC. To substantiate this, we analyze the runtime as a function of $L$ in Fig.~\ref{pilotSequences}(c).  From this figure, it is evident that AEM-ADMM and AEM-SBL are more scalable than their classical counterparts. For example, the AEM-SBL runs five times faster than the SBL as $L$ increases. Although Table \ref{complexAnalysis} indicates that both AEM-ADMM and ADMM's runtimes are insensitive to $L$, it is worth mentioning that the former exhibits nearly 10 times faster runtime than the latter as $L$ increases, attributed to its faster convergence. Despite its faster runtime compared to SOMP, CB-SOMP generally produces poor results, as demonstrated in Figure \ref{pilotSequences}(a) and \ref{pilotSequences}(b). \textcolor{black}{Interestingly, even though poorly performs under the current setting, the VAMP is highly scalable, which is consistent with the results in \cite{senel2018grant,cheng2020orthogonal,rangan2019vector,donoho2009message}.}
% \textcolor{red}{
%  The reason for this improvement is that longer pilot sequences within each cluster guarantee fewer collisions and thus increasing the reliability of the detection, through improved RIP of the matrix $\bm{A}_g$. The AEM-SBL perform similar to its classical counterpart thus proving the reliability of the proposed algorithm. However, both AEM-ADMM and ADMM have comparable performance in very small pilot ratios, e.g., at approximately $\frac{L}{N}= 0.12$, both AEM-ADMM and ADMM achieve the NMSE of $0.1$. Given the massiveness of the MTDs in an MTC network, the lower ratios are of interest due to their practicality. On the other hand, the naive application of the SOMP, i.e., CB-SOMP leads to poor channel estimation capabilities in general as compared to the classical SOMP. This is mainly due to the fact that the naive application relies on a completely mismatched model, which justifies the need for the AEM in the developed algorithms. Even though high ratios are not practical, it can be observed that SOMP performs significantly well in high $\frac{L}{N}$. This can be attributed to the fact that it relies on the orthogonal pilot sequences, which are encouraged by longer $L$. Despite this, SOMP and ADMM do not consider the presence of clusters. }

{\color{black} In Fig.~\ref{pilotSequencesVsChannel}, we analyze the impact of the proposed pilot sequences on the performance of JADCE when using AEM-SBL, considering both correlated and uncorrelated Rayleigh fading channels. Generally, we observe that the performance improves with increasing pilot lengths, although it eventually reaches a saturation point under the correlated channel model. Notably, even though both Gaussian and Bernoulli matrices are not orthogonal basis, they substantially improve the performance of JADCE as $L$ increases. In spite of not being an orthogonal basis, longer lengths of the Gaussian and the Bernoulli sensing pilot sequences improve the approximation of $\hat{\mathbf{Y}}$, which improves the detection capabilities of the AEM-SBL. Consistent with the results of \cite{senel2018grant}, the Bernoulli pilot sequences have better capabilities in facilitating JADCE. However, it is more practical to have very low pilot-to-device ratios for scenarios with a massive number of devices transmitting short packets, such as MTDs. In view of this, the proposed cluster-based pilot performs better than the other pilot sequences in a practical setting. For instance, observe that with the proposed pilot sequences, it is possible to get a PMD of less than $0.001$, while Bernoulli-based pilot sequences and the Gaussian both achieve a PMD greater than $0.01$ for the pilot-to-device ratio of less than $0.1$. In general, the AEM-SBL portrays the same performance trend for correlated and uncorrelated channels, with the only difference being that the performance is generally poorer under strong spatially correlated channels, such as the case of $\sigma_{\phi_{n,g}} = 5^{\circ}$. However, this is an expected phenomenon because correlated channels reduce the channel hardening \cite{bjornson2017massive}.
\begin{figure}[t!]
\centering \includegraphics[width=0.475\textwidth]{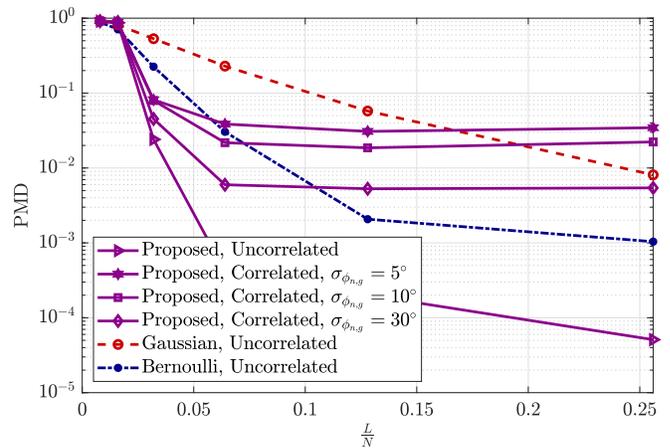}
\caption{\textcolor{black}{Performance as a function of the pilot length for uncorrelated channels and spatially correlated channels with $\sigma_{\phi_{n,g}} =\{ 5^{\circ}, 10^{\circ}, 30^{\circ}\}$ using AEM-SBL.}}
\label{pilotSequencesVsChannel}%\vspace{-7mm}
\end{figure}}
\subsection{On the average SNR}
\begin{figure}
\centering  
     \includegraphics[width=0.475\textwidth]{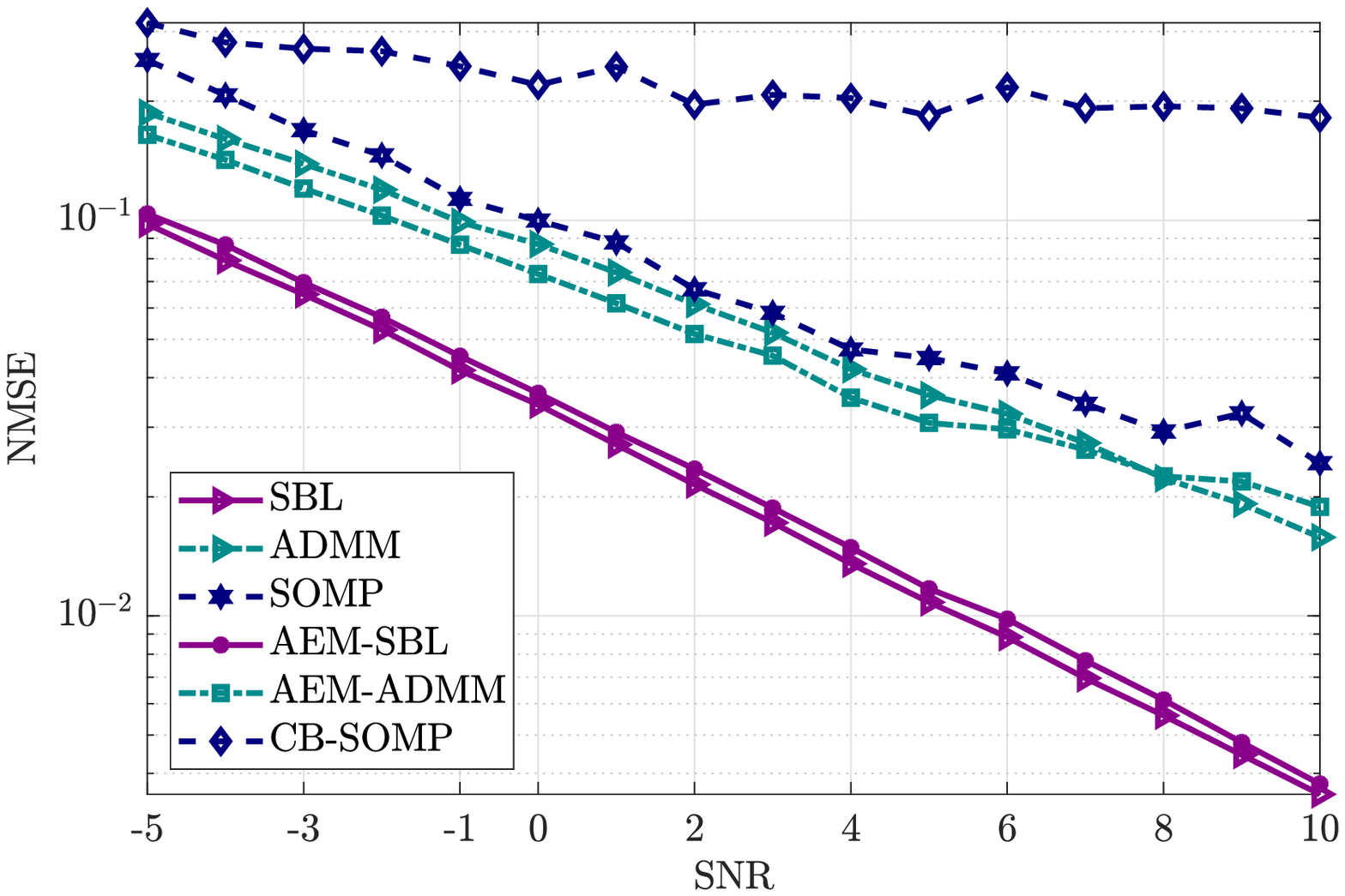}\\
 \includegraphics[width=0.475\textwidth]{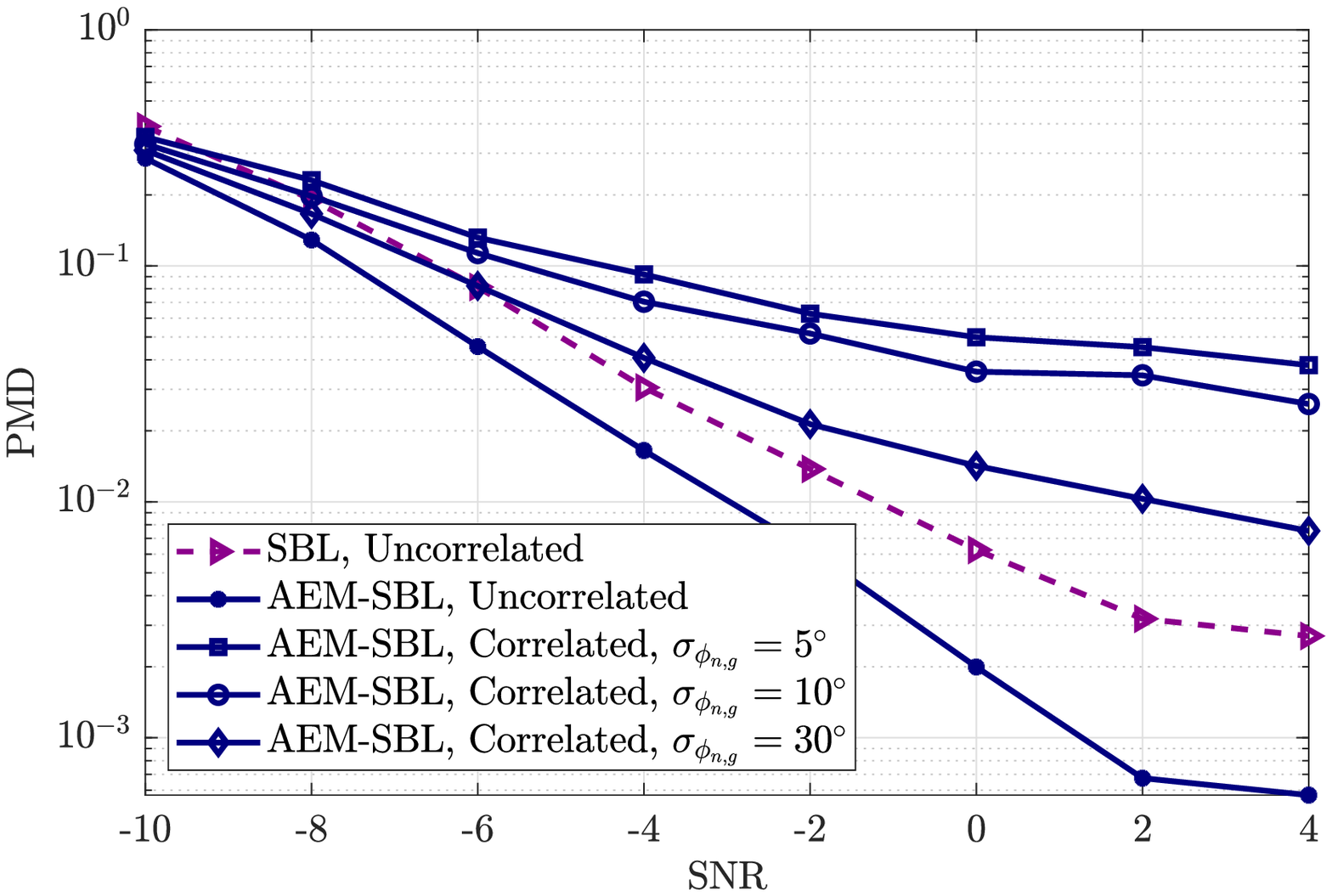}
\caption{Performance in terms of a) NMSE \textcolor{black}{(top), b)  PMD (bottom) as a function of the SNR for uncorrelated and spatially correlated channels with $\sigma_{\phi_{n,g}} \in\{ 5^{\circ}, 10^{\circ}, 30^{\circ}\}$ using AEM-SBL.}}
\vspace{-3mm}
\label{SNRfigures}
\end{figure}

\begin{figure}[t!]
\centering
%\begin{subfigure}{0.5\textwidth}   
\includegraphics[width=0.475\textwidth]{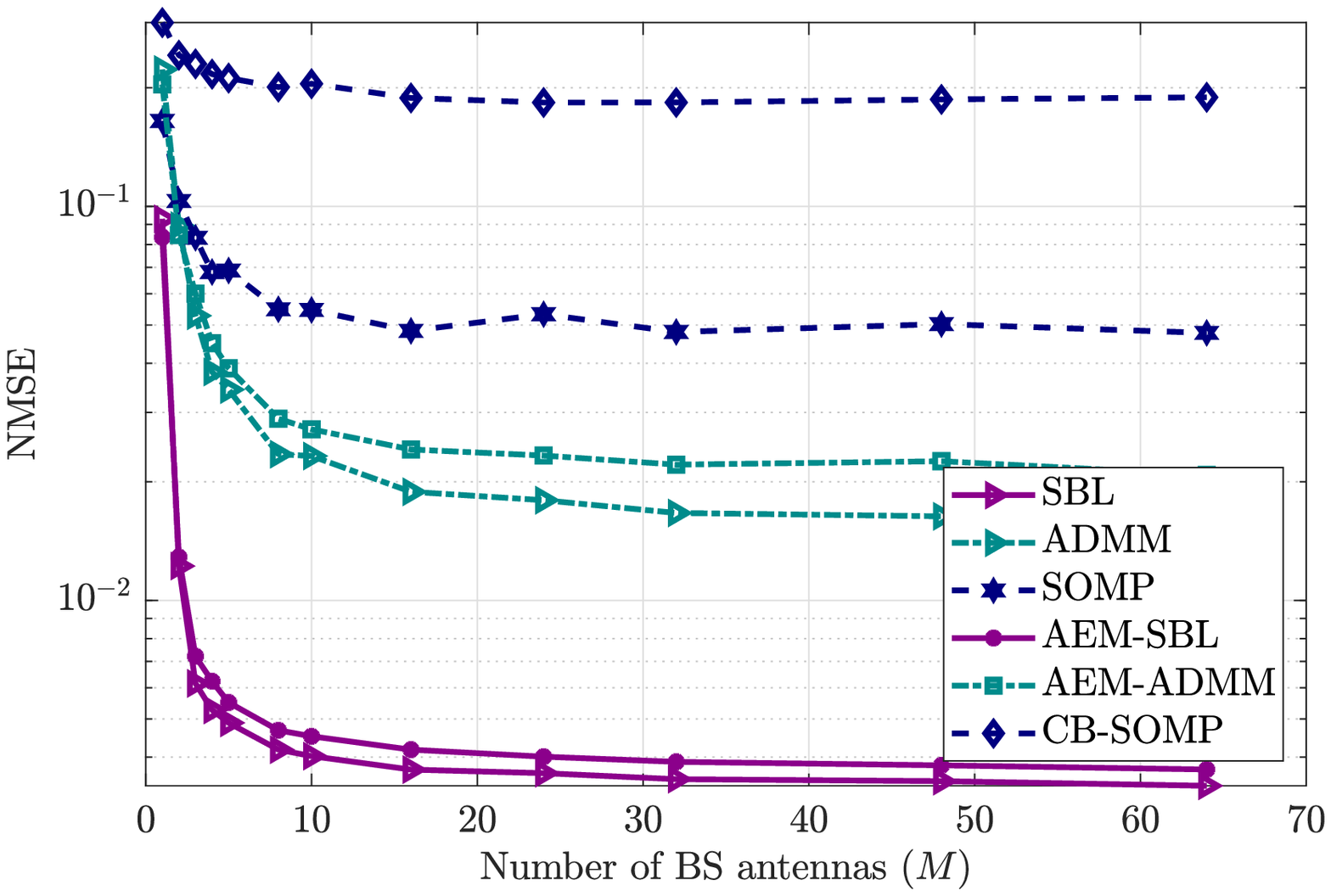}\\
 % \caption{~}
  %  \label{antennaNMSE}
%\end{subfigure}
%\hfill
%\begin{subfigure}{0.5\textwidth}    
\includegraphics[width=0.475\textwidth]{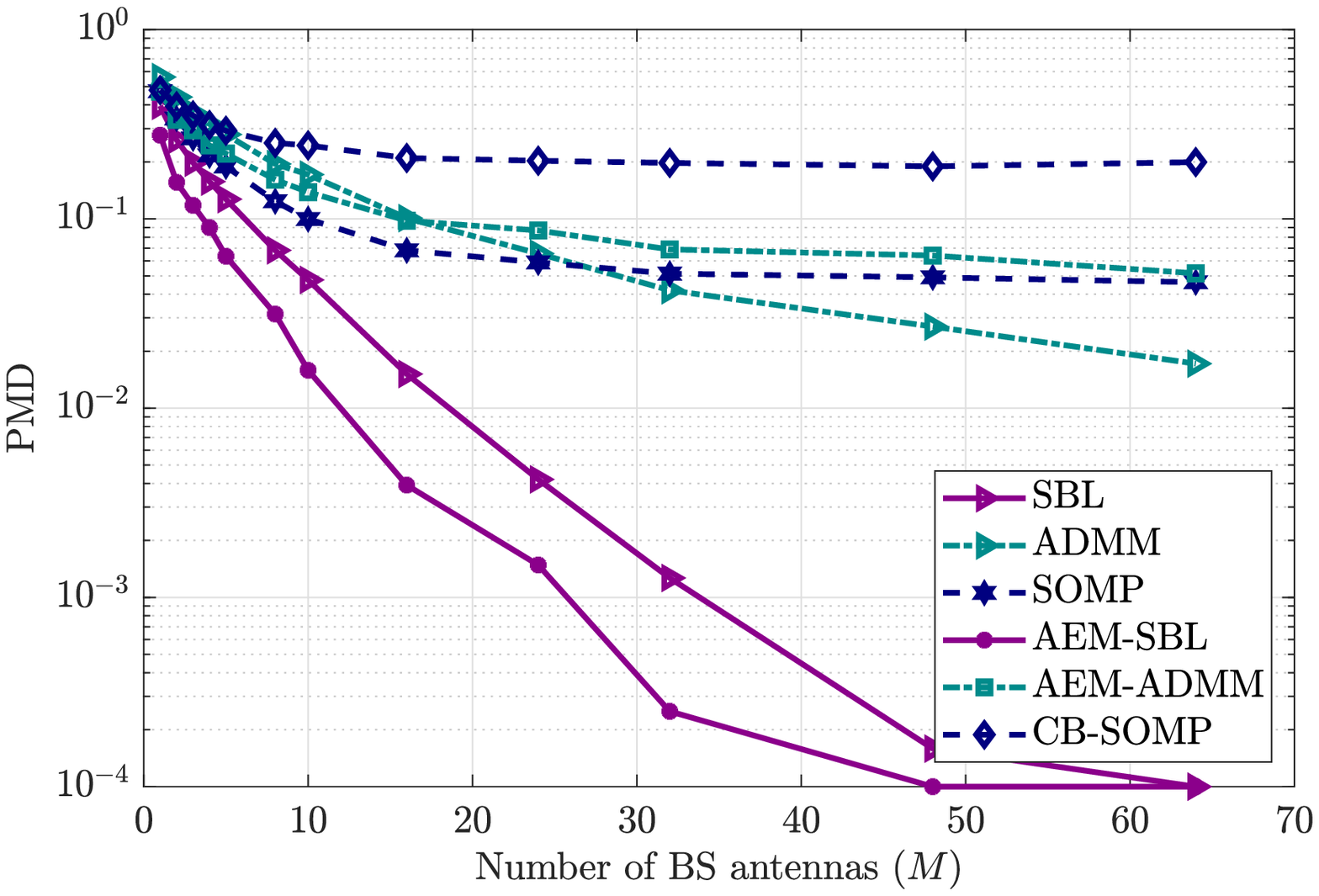}\\
%\caption{~}
 %  \label{antennaPMD}
%\end{subfigure}
%\hfill
%\begin{subfigure}{0.5\textwidth}   
\includegraphics[width=0.475\textwidth]{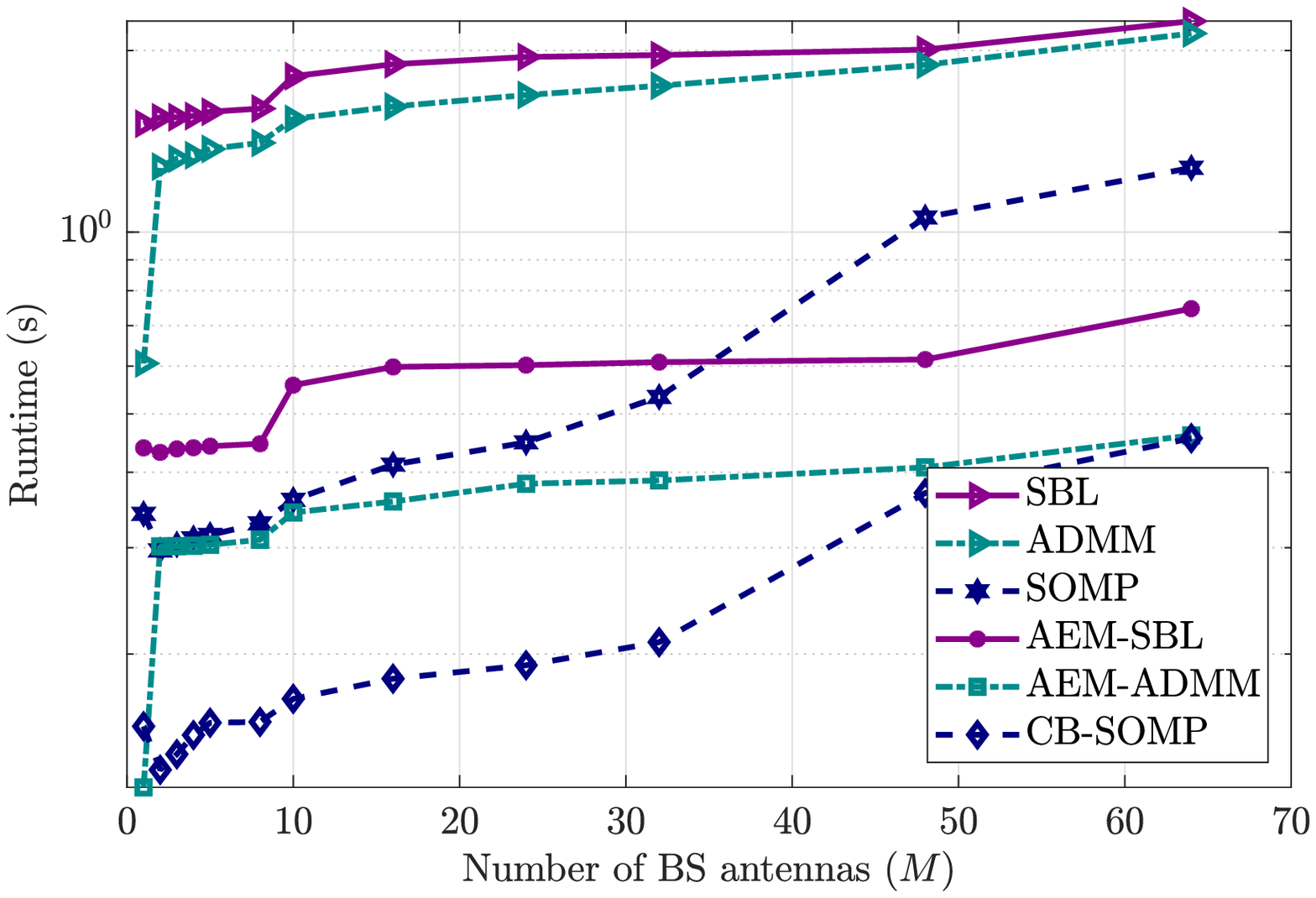}\\
%\caption{~}    
%\label{RuntimeAntennas}
%\end{subfigure}
\caption{Performance in terms of a) NMSE \textcolor{black}{(top), b) PMD (middle)}, and c) run-time \textcolor{black}{(bottom)} as a function of the number of antennas.}
\vspace{-3mm}
\label{antennaPerformance}%\vspace{-7mm}
\end{figure}

In Fig.~\ref{SNRfigures}, we present the \textcolor{black}{performance results} as a function of the average SNR. Specifically, Fig. \ref{SNRfigures}(a) illustrates the performance in terms of the NMSE. As shown, the performance of all the JADCE algorithms improves as the signal power gets higher than the noise power, i.e., increasing average SNR. In spite of this, CB-SOMP doesn't improve in performance due to its reliance on the mismatched model. In view of this, increasing SNR increases the amount of the mismatch, thus resulting in inferior performance than other algorithms. For both AEM-SBL and AEM-ADMM, the AEM tends to correct the mismatch more accurately with increased power levels, and this results in their superior performance at high SNR. \textcolor{black}{On the other hand, Fig.~\ref{SNRfigures}(b) shows the performance of AEM-SBL and SBL in terms of the PMD under the uncorrelated Raleigh fading channel. In addition, we evaluate the performance of AEM-SBL for varying degrees of spatial correlation. Generally, both algorithms demonstrate improved detection capabilities, reflected by low PMD as the SNR increases. From the figure, AEM-SBL exhibits superior detection capability compared to SBL, despite Fig.~\ref{SNRfigures}(a) showing slightly better NMSE for SBL. However, it is important to note that the performance of PMD depends on the threshold setting, which is set to obtain $\text{PFA} = 10^{-3}$ in the sequel. Therefore, we cannot claim the algorithmic superiority of AEM-SBL solely based on this threshold. Nonetheless, both AEM-SBL and SBL demonstrate comparable performances, while AEM-SBL has shorter run-times. However, as the correlation in the channel increases, i.e., $\sigma_{\phi_{n,g}}$ decreases, the performance of AEM-SBL is degraded due to the loss of channel hardening.}
\subsection{On the number of antennas in the BS}
In Fig~\ref{antennaPerformance}, we present the \textcolor{black}{performance results} as a function of the number of antennas in the BS. \textcolor{black}{The} overall trend of Fig.~\ref{antennaPerformance}(a) \textcolor{black}{indicates} that all the algorithms improve the channel estimation accuracy as the number of antennas at the BS increases. This is mainly \textcolor{black}{because} the increase in $M$ provides an additional structure that can be exploited during the signal recovery process\cite{senel2018grant,wei2018joint}.  However, the improvements get to saturation as shown by no substantial improvement beyond $M = 20$, which is consistent with the results of \cite{senel2018grant}. We also note that both the AEM-SBL and SBL outperform the other algorithms due to their Bayesian nature. Similarly, in Fig.~\ref{antennaPerformance}(b), the performance in terms of the PMD show an improvement as the number of antenna increases and this is due to the high resolution of the MMV problem under a large number of antennas\cite{zheng2021jointMMV}. However, it is important to note that increasing $M$ has a substantial impact on the run-time and scalability of the algorithms as illustrated in Fig.~\ref{antennaPerformance}(c), where it is shown that the run-time significantly increases with the number of antennas. This is due to the increasing matrix dimensions. In spite of this, it is evident that both AEM-SBL and AEM-ADMM run faster than their conventional counterparts, thus they may facilitate network scalability. This is one of the major benefits of the proposed algorithms, which exploit the structure of the sensing matrix.  

\subsection{On the activation probability}
In Fig.~\ref{sparseFunction}, we analyze the impact of the activation probability on the JADCE performance of the algorithms by evaluating NMSE. \textcolor{black}{It} can be observed that all the algorithms perform poorly with decreasing sparsity level, i.e., as more devices are activated at the same time (high $\epsilon$). This is due to the fact that there is an increase in MUI in each cluster when the sparsity is decreased. Notably, the AEM-based algorithms demonstrate comparable efficiency to their counterparts by efficiently reducing the MUI, while handling JADCE problems of reduced dimensions.
\begin{figure}[t]
    \centering    \includegraphics[width=0.475\textwidth]{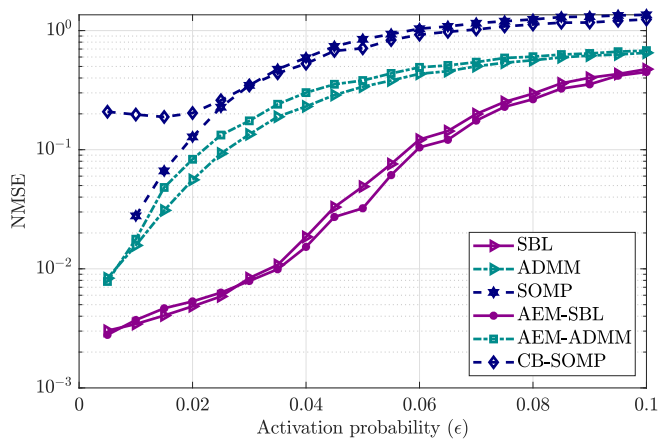}
\caption{NMSE as a function of $\epsilon$.}
    \label{sparseFunction}
\end{figure}

 {\color{black}\subsection{On the cluster sizes}
 Fig. \ref{mtdClusters} shows the performance in terms of the run-time as a function of the number of MTDs and clusters. Results show that the run-time increases with the number of devices. For instance, with $N = 64$, SBL runs for less than $0.1$ seconds, while $N = 8192$ takes approximately $100$ seconds. This agrees and corroborates the complexity analysis in Table.~\ref{complexAnalysis}. However, it can be observed that both AEM-SBL and AEM-ADMM have fewer run times than the centralized SBL and ADMM as the number of clusters increases. These results demonstrate the scalability of the proposed JADCE solutions.}
% \textcolor{black}{Fig~\ref{AER} shows performance in terms of the AER as a function of $\epsilon$. In general, the AER increases with the increase in sparsity. This is mainly due to the increase in the MUI among the devices. In spite of that, the proposed algorithms perform equally to the convectional ones, showing their success. At practical sparsity levels, e.g., $\epsilon = 0.01$ AEM-ADMM performs similar to ADMM´and CB-SOMP, while SBL, AEM-SBL and perform similar to SOMP. }

% \begin{figure}[t]
%    \centering    \includegraphics[width=0.45\textwidth]{resultsFigs/iterationFigure.eps}
% \caption{NMSE as a function of the iteration steps}  \label{ConvergenceAnlalysis}
% \end{figure}
% \textcolor{black}{Fig~\ref{ConvergenceAnlalysis} presents the convergence analysis. From the results, it can be seen that AEM-SBL converges at a similar rate as SBL. However, both algorithms converge to $10^{-4}$ in NMSE. On the other hand, the ADMM and AEM-SBL converge with lower NMSE. It is however notable that AEM-SBL is slower in convergence than ADMM. This is consistent with works such as. In spite of this, the run-time for the AEM-SBL and AEM-ADMM is drastically reduced by the clustering framework as presented in Fig.~\ref{RuntimePilots}. }

 \begin{figure}[t]
     \centering     \includegraphics[width=0.475\textwidth]{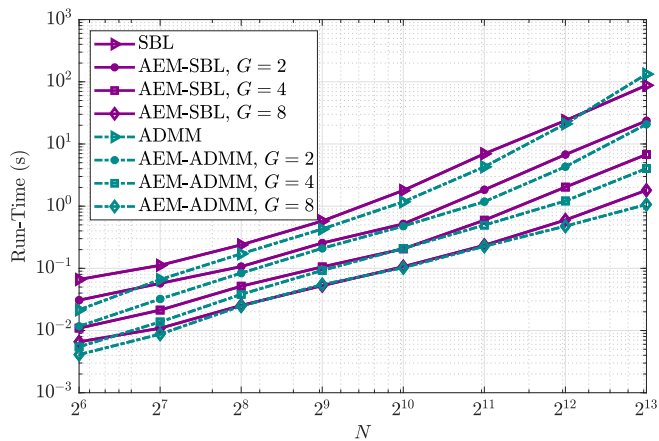}
     \caption{\color{black}Performance as a function of the number of MTDs and the cluster sizes.}
     \label{mtdClusters}
 \end{figure}  

\section{Conclusion and future works}
\label{Conclude}
This work presented a novel framework for cluster-based device activity detection and channel estimation that relies on orthogonal pilot subspaces to optimize GF-NOMA. By utilizing non-i.d.d. pilot sequences, the proposed pilot-based clustering approach promotes efficient device activity detection and enhances network flexibility and scalability, making it practically relevant. We leveraged concepts from the field of inverse problems, and we proposed novel data-driven JADCE solutions: i) AEM-ADMM, which uses iterative soft thresholding for scenarios without exact priors, and ii) AEM-SBL, designed for cases where prior distributions can be formulated.
The proposed algorithms \textcolor{black}{outperform} their classical counterparts \textcolor{black}{in terms of} run-time while maintaining similar performance.
Furthermore, the AEM introduces a fresh perspective on device active detection in MTC by accounting for the impairments of sensing matrices. \textcolor{black}{Our proposal constitutes} a timely solution for receivers in a 5GB cellular network. \par
As a potential avenue for future research, the AEM approach presented in this work can be further enhanced by introducing adaptive correction parameters at each iteration, which may improve its performance. Additionally, the flat fading channel assumption can be relaxed to frequency selective fading to address  orthogonal frequency-division multiplexing (OFDM)-inspired mMTC. Furthermore, the results of this work can be extended to cell-free MIMO communication systems. }  

% Another interesting research direction would be to extend the present work to cell-free scenarios.  

\ifCLASSOPTIONcaptionsoff
  \newpage
\fi

\bibliographystyle{IEEEtran} 
%\bibliography{ref.bib}
 %%COMPRESSION TWEAK
\bibliography{jour_short,conf_short,marataManuscript.bbl}

%\clearpage

\printglossary[type=\acronymtype]

\printglossary
\end{document}